\documentclass[12pt]{article}
\pdfoutput=1
\usepackage{amsmath,amssymb}
\usepackage{graphicx}
\usepackage{cite,fancyhdr}
\usepackage{bm, color}
\usepackage{amssymb,amsfonts,slashed,amsthm,amsmath,graphicx,soul,empheq}
\usepackage[caption=false]{subfig}
\usepackage{hyperref}
\newcommand{\Slash}[1]{{\ooalign{\hfil#1\hfil\crcr\raise.167ex\hbox{/}}}}

\newcommand{\beq}{\begin{equation}}  \newcommand{\eeq}{\end{equation}}
\newcommand{\bef}{\begin{figure}}  \newcommand{\eef}{\end{figure}}
\newcommand{\bec}{\begin{center}}  \newcommand{\eec}{\end{center}}
  
\newcommand{\laq}[1]{\label{eq:#1}}  

\newcommand{\Eq}[1]{Eq.(\ref{eq:#1})}

\newcommand{\eq}[1]{(\ref{eq:#1})}
\newcommand{\Sec}[1]{Sec.\ref{chap:#1}}
\newcommand{\ab}[1]{\left|{#1}\right|}
\newcommand{\vev}[1]{ \left\langle {#1} \right\rangle }

\newcommand{\lac}[1]{\label{chap:#1}}
\newcommand{\SU}[1]{{\rm SU{#1} } }

\def\({\left(}
\def\){\right)}

\def\O{\mathcal{O}}
\def\U{\mathop{\rm U}}

\newcommand{\AND}{~{\rm and}~}
\newcommand{\EV}{ {\rm eV} }
\newcommand{\KEV}{ {\rm keV} }
\newcommand{\MEV}{ {\rm MeV} }
\newcommand{\GEV}{ {\rm GeV} }
\newcommand{\TEV}{ {\rm TeV} }

\def\d{\delta}

\def\l{\lambda}
\def\m{\mu}

\def\x{\xi}

\def\D{\Delta}
\def\G{\Gamma}

\def\L{\Lambda}
\def\F{\Phi}
\def\P{\Psi}

\def\*{\dagger}

\usepackage[affil-it]{authblk}
\usepackage[left=2cm,right=2cm,top=2cm,bottom=2cm,a4paper]{geometry}

\begin{document}
\begin{titlepage}
\begin{center}
\setcounter{footnote}{0}
\setcounter{figure}{0}
\setcounter{table}{0}

\hspace{3cm}

\hspace{3cm}

\hspace{3cm}

\hspace{3cm}

{\Large\bf 
Feebly-Interacting Peccei-Quinn Model}

\vskip .75in

{ \large    Wen Yin}

\vskip 0.25in

\begin{tabular}{ll}
& {\em Department of Physics, Tokyo Metropolitan University,} \\
&{
Minami-Osawa, Hachioji-shi, Tokyo 192-0397 Japan} \\[.3em]
&\!\! {\em Department of Physics, Tohoku University, }\\
& {\em Sendai, Miyagi 980-8578, Japan}\\[.3em]
&\\
& 
\end{tabular}

\begin{abstract}
The QCD axion is widely studied as a dark matter (DM) candidate and as a solution to the strong CP problem of the Standard Model. In conventional field-theoretic models, a much larger mass scale than the electroweak (EW) scale is typically introduced to spontaneously break Peccei-Quinn (PQ) symmetry with a large enough axion decay constant, $f_a$, thereby avoiding constraints from star cooling. In this paper, I propose an alternative approach to achieving the large decay constant: a PQ scalar field with a large wave function renormalization constant, analogous to a feebly coupled gauge theory. Other dimensionless parameters are $\mathcal{O}(1)$ in the unit of the EW scale for the naturalness. This framework predicts a light PQ Higgs boson with a mass $\sim (\mathrm{EW~scale})^2 / f_a$. Exotic particles associated with the PQ anomaly are expected to have masses around the EW scale. The proposed model alleviates both the PQ quality and EW scale fine-tuning problems and introduces interesting axion-PQ Higgs cosmologies, encompassing: slim axion DM from a fat string network, heavy axion DM from PQ Higgs condensate fragmentation, PQ Higgs DM, and axion-PQ Higgs co-DM scenarios. 
Potential experimental signatures are explored, including fifth-force tests, DM detections,  accelerator searches, and gravitational wave observations by employing lattice simulation. Possible extensions of the scenario are also discussed.
\noindent
\end{abstract}

\end{center}
\end{titlepage}
\setcounter{footnote}{0}
\setcounter{page}{1}

\section{Introduction and Setup}

In the Standard Model, there is a fine-tuning problem associated with the strong CP phase, as inferred from the non-observation of the neutron electric dipole moment and the non-vanishing quark masses~\cite{ParticleDataGroup:2020ssz}. The QCD axion is a promising solution to the strong CP problem~\cite{Peccei:1977hh,Peccei:1977ur,Weinberg:1977ma,Wilczek:1977pj}, and it is one of the leading candidates for dark matter (DM).

In fact, in minimal setups, the axion may have another fine-tuning problem due to the large difference between the decay constant, which lies in the range~\cite{Mayle:1987as, Raffelt:1987yt,Chang:2018rso}\footnote{There is a conventional upper bound from the axion dark matter abundance~\cite{Preskill:1982cy,Abbott:1982af,Dine:1982ah}. This bound can be absent if the Hubble parameter during inflation, which lasts long enough, is lower than the QCD scale~\cite{Graham:2018jyp, Guth:2018hsa}.} \beq \laq{window} 10^{8}\GEV \lesssim f_a, \eeq and the electroweak (EW) scale $v_{\rm EW}\sim 170\GEV$. 
Usually, we introduce a Peccei-Quinn (PQ) field $\F$ whose vacuum expectation value $\vev{\F}\sim f_a/\sqrt{2}$ breaks the PQ symmetry $\U(1)_{\rm PQ}$, and very na\"{i}vely, this implies the hierarchy between the PQ Higgs boson mass $f_a$, and the EW scale.

For concreteness, I consider the KSVZ model in this paper~\cite{Kim:1979if, Shifman:1979if}, although our discussion also applies to the DFSZ model~\cite{Dine:1981rt, Zhitnitsky:1980tq} (see \Sec{cd}). Then, the renormalizable Higgs potential is \beq \laq{pot} V\supset \lambda_P |\F|^2 | h|^2 - \mu_h^2 | h|^2 - m_\F^2 | \F|^2 + \lambda_h | h|^4 + \lambda | \F|^4, \eeq where $h$ is the Standard Model Higgs doublet field.
We introduced the portal coupling $\lambda_P\neq 0$ because no symmetry forbids it. Even if we set $\lambda_P=0$ at the tree level, it is generated at the loop level since $\F$ has to couple to PQ quarks to induce the chiral anomaly to $\SU(3)c$ to solve the strong CP problem; e.g., \beq {\cal L}_{\rm int} = -y \F \bar{\P} P_L \P + h.c. \eeq Here, we introduced a pair of chiral PQ quarks $\P$, which are charged under the PQ symmetry. $\P$ and $\bar \P$ are (anti-)fundamental representations of $\SU(3)_c$, so the domain wall number is $1$, thereby evading a domain wall problem.
One finds from \Eq{pot} that a correction to the Higgs field mass squared, of order $\lambda_P f_a^2$, exists. This implies a serious fine-tuning to realize the EW scale if $\lambda_P \sim 1$. One way to solve the fine-tuning problem is to introduce new particles via supersymmetry. Alternatively, it may be possible that the EW scale is special from the viewpoint of the anthropic principle, and explaining it through model building is not necessary.

In this paper, I explore a new possibility motivated by the fine-tuning problem. To explain this, let us consider the kinetic term of the PQ field, written as 
\beq 
{\cal L} = Z |\partial \Phi |^2, 
\eeq 
where $Z$ is the wave-function renormalization constant of the field $\Phi$. 
The proposed {\it feebly interacting PQ (FIPQ) model} is defined such that, in a basis obtained via field redefinitions, all introduced parameters are of $\mathcal{O}(1)$ in units of a single mass scale $\Lambda$, while $Z$ is much larger than unity. Furthermore, I assume that $\Lambda$ is around the EW scale to avoid introducing an additional hierarchy and thus fine-tuning. In particular, the mass parameter for $\Phi$ in this basis $ \sim \Lambda$, which is also around the EW scale. 
Note that $\Lambda$ does not need to be the cutoff scale of the theory, as no perturbative unitarity is violated in this framework.
 The FIPQ model resembles the weakly coupled limit of a gauge theory, where one can switch to a basis in which the wave function renormalization constant of the gauge field is inversely proportional to the square of the small gauge coupling and therefore, becomes large. 
 
 The number of extremely large parameters in the FIPQ model is the same as in the original PQ model:
\beq
\laq{conp}
\frac{m_\Phi^2}{v_{\rm EW}^2} \gg 1 \quad [\text{PQ model}], \qquad\qquad Z \gg 1 \quad [\text{FIPQ model}],
\eeq
and the magnitude of the large quantities is also comparable. The advantage of the FIPQ model is that it avoids the aforementioned fine-tuning problem related to the EW scale, and it alleviates the PQ symmetry quality problem caused by small instantons. Some ultraviolet (UV) completions can lead to this setup,\footnote{\label{ft1} One UV realization of this setup involves a massless and free complex field, $\Phi$, residing in the 5D bulk of an extra-dimensional model with the symmetry $\Phi \to \Phi + \alpha$, where $\alpha$ is an arbitrary complex number, while all other fields are localized on a 4D brane. In this scenario, $Z \sim L \Lambda_{\rm 5D}$ arises in the low-energy 4D theory after compactification. If the compactification radius $L$ is much larger than the 5D cutoff scale $1/\Lambda_{\rm 5D}$ in the 5D kinetic term of $\F$, a sizable wave function at the compactification scale can be achieved.} although I perform a UV model irrelevant discussion in this paper.
 
In this paper, I also show that very non-trivial particle spectra and DM cosmology emerge in this model. In general, the PQ Higgs boson is predicted to be much lighter than the EW scale, and the particles coupled to the PQ Higgs boson to induce the chiral anomaly are around the EW scale.
In particular, the minimal realization of the FIPQ proposal yields three distinct DM scenarios depending on the reheating temperature and coupling, especially $\lambda_P$: 
(IIa) slim axion DM from a fat string network~\cite{Saikawa:2024bta,Kim:2024wku,Buschmann:2024bfj,Benabou:2024msj,Kim:2024dtq} (see also \cite{Sikivie:1982qv,Vilenkin:1982ks,Harari:1987ht,Davis:1986xc}, and \cite{Dine:2020pds,Saikawa:2024bta} for discussions),
(Ib) axion DM from the fragmentation of the PQ Higgs condensate, and 
(IIb) PQ Higgs DM from thermal misalignment~\cite{Batell:2021ofv,Batell:2022qvr} during reheating and/or axion DM from vacuum/stochastic misalignment~\cite{Preskill:1982cy,Abbott:1982af,Dine:1982ah,Graham:2018jyp, Guth:2018hsa}.

The scenario (IIa) is well-known, but in our case, due to the very light PQ Higgs, the string is fat, modifying the prediction for the axion mass by a factor of 2. The  scenario (Ib), specific to this model, has not been noticed so far and predicts very strong gravitational waves, and perhaps miniclusters. 
The PQ Higgs may be probed in accelerators. The relevant dynamics is non-linear and I use lattice simultion to clarify it. 
In (IIb), the parameter region involves the one consistent with the multicomponent DM consisting of both the axion and the PQ Higgs.

A related context may be the clockwork scenario where the axion becomes weakly/strongly coupled by introducing many axions mixing with each other in certain manner~\cite{Choi:2015fiu,Kaplan:2015fuy,Giudice:2016yja,Higaki:2016yqk}, which can alleviate the fine-tuning problem and the quality problem~\cite{Higaki:2016yqk}. In my proposal,  rather than the axion, it is the PQ Higgs field that is feebly coupled, which leads to the novel PQ Higgs-axion cosmology.

This paper is organized as follows. In \Sec{1}, I discuss the spectrum of the FIPQ model and show the parameter region for the light PQ Higgs, and the alleviation of the quality problem and the EW scale fine-tuning problem in the canonical basis. In \Sec{2}, I provide a detailed discussion on DM production in the context of inflationary cosmology. After a brief review of the inflationary dynamics of the Higgs field and the thermal potential in \Sec{2-1} and \Sec{2-2}, I analyze slim axion DM from a fat string network in \Sec{2-3} and \Sec{2-4}, heavy axion DM from PQ Higgs condensate fragmentation in \Sec{2-5}, and PQ Higgs DM as well as axion-PQ Higgs co-DM scenarios in \Sec{2-6}.  Finally, the last section, \Sec{cd}, is devoted to conclusions and discussions, where I explore extensions of the model and possible novel mechanisms related to it.

\section{Masses and Couplings in the Canonical Basis and Quality Problem}
\lac{1}
\paragraph{Spectra and couplings in the canonical basis}
In this section and in the following, I will write the parameters only in a canonically normalized basis. By normalizing the kinetic term, I immediately obtain \begin{align} m_\Phi \sim \frac{ \Lambda }{\sqrt{Z}}, \lambda_P \sim \frac{1}{Z}, y\sim \frac{1}{\sqrt{Z}}, \lambda \sim \frac{1}{Z^2}.\laq{cupnom} \end{align} 
The symbol $\sim$ indicates that they do not differ by several orders of magnitude.
 Other parameters do not change. 

As we will see, to alleviate the fine-tuning problem discussed in the introduction, 
the EW scale and the typical mass scale of the model, $\Lambda$, should not be too different; say, \beq \Lambda \sim \TEV. \eeq 
Therefore, we get the axion decay constant, \beq f_a = \sqrt{2}\vev{\Phi}\sim \sqrt{Z}\Lambda \sim \sqrt{Z} \TEV. \eeq Thus, by requiring \Eq{window}, we obtain $ \sqrt{Z} \gtrsim 10^{5}. $ 
The robust prediction of the model is a very weakly coupled, very light PQ Higgs boson, $s$, appearing in the decomposition $\Re\Phi= \vev{\Phi}+\frac{s}{\sqrt{2}}$. The mass is given by \beq
\laq{masss} m_s\simeq \sqrt{2}m_\Phi\sim \frac{\Lambda^2}{f_a}\sim \frac{1}{\sqrt{Z}}\Lambda. \eeq To satisfy the supernova bound on $f_a$, \Eq{window}, we get \beq m_s \lesssim 10^{-5}\Lambda\sim 10\MEV. \eeq There is an interesting relation between $m_s$ and the axion mass $m_a\simeq \sqrt{\chi_0}/f_a$, which is \beq \frac{m_a}{m_s}\sim \frac{\sqrt{\chi_0}}{\Lambda^2}\sim 10^{-8}, \eeq independent of the decay constant $f_a$, where $\chi_0\approx (0.08\GEV)^4$ is the topological susceptibility.

The PQ fermion is not light because \beq m_\Psi\sim y \vev{\Phi} \sim \Lambda, \eeq which does not depend on $Z$. Thus, this fermion can be an experimental target for collider searches. For instance, it can be pair-produced and decay into a Standard Model quark and standard gauge/Higgs boson via mixing assuming it has a similar representation as a Standard Model quark.
 This process may constrain $m_{\Psi}\gtrsim 2\TEV$~\cite{CMS:2019afi,ATLAS:2022ozf}, depending on the coupling, which is naturally $\O(1)$. 
 This is also a constraint for $\L$ not less than $\O(\TEV)$.

\paragraph{Alleviation of the EW scale fine-tuning}
An advantage of the FIPQ model is its ability to alleviate the EW scale fine-tuning problem compared to the original PQ model, despite both models involving unnaturally large parameters in different parts (\Eq{conp}). 

To illustrate this, consider the correction to the Standard Model-like Higgs mass arising from the portal coupling:
\beq 
\Delta \mu_h^2 \sim -\lambda_P \vev{|\Phi|}^2 = \mathcal{O}(\Lambda^2),
\eeq
which represents the correction to the Higgs mass parameter. The loop-induced contribution from $\Psi$ is further suppressed by loop factors since $\Psi$ has a mass of $\mathcal{O}(\Lambda)$. The contributions from $s$ and $a$ loops to the EW scale are even more suppressed due to their small masses and couplings. Consequently, if $\Lambda$ is around the weak scale, the EW scale remains stable under these corrections in this scenario.

\begin{figure}[!t] 
\begin{center}
\includegraphics[width=145mm]{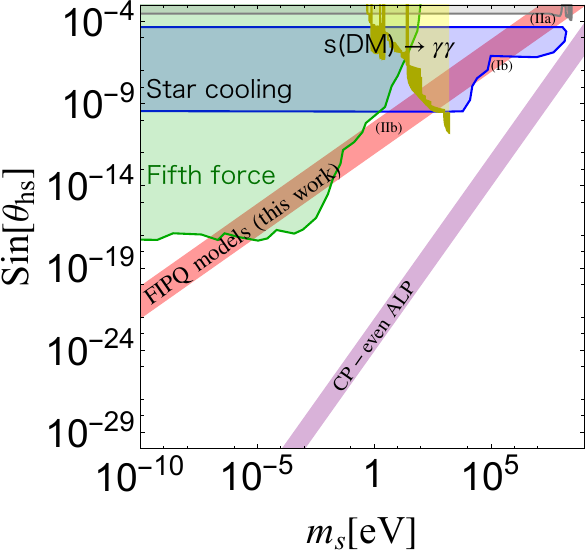}
\end{center} 
\caption{The prediction is shown in the $s$ mass and Higgs mixing parameter plane (red band). Constraints from the fifth forces (green-shaded region), star cooling (blue-shaded region), and accelerator searches (gray-shaded region) are also indicated. Assuming $s$ as the dominant DM, the excluded region from DM decaying into two photons is shown in the yellow-shaded region. Additionally, we present the prediction of the CP-even axion model (purple band), which also predicts a typical relation between the mass and the Higgs mixing, for comparison~\cite{Sakurai:2021ipp,Haghighat:2022qyh}. The labels (Ib), (IIa), and (IIb) correspond to the dark matter cosmological scenarios discussed in the next section.
The accelerator searches and star cooling are adopted from \cite{Sakurai:2021ipp}, and DM decay constraints from \cite{AxionLimits}.
}
\label{fig:1} 
\end{figure}

It is also worth noting that the large $Z$ (or the small couplings and masses) discussed here are natural in the sense of the 't Hooft criterion (within quantum field theory)~\cite{tHooft:1979rat}. In the limit $Z \to \infty$, $\Phi$ becomes a free particle with an infinite number of symmetries.\footnote{Requiring the coupling to be stronger than gravity, $y \sim 1/\sqrt{Z} > m_\Phi / M_{\rm pl} \sim \Lambda / M_{\rm pl}$, we find $f_a \sim \Lambda \sqrt{Z} \lesssim M_{\rm pl}$. While Fig.~\ref{fig:1} illustrates the parameter region for super-Planckian decay constants, this paper only considers the case $f_a \lesssim M_{\rm pl}$.}

\paragraph{Parameter region}
The light PQ Higgs boson is very weakly coupled to the PQ fermions as well as to the Standard Model-like Higgs doublet due to the suppressed coupling \eq{cupnom}. Since the PQ fermions and Standard Model-like Higgs couple to the other Standard Model particles, feeble interactions with the Standard Model particles are induced. For concreteness, let us neglect the PQ fermion loop/mixing contribution, which depends on the charge assignment and flavor structure of the PQ fermion. Then, the dominant contribution comes from the Higgs mixing \beq V\supset \lambda_P \vev{\Phi} v_{\rm EW} \frac{h s}{2} \sim \frac{\Lambda}{\sqrt{Z}} v_{\rm EW} h s. \eeq Then we can easily obtain the mixing parameter \beq \theta_{hs} \sim \frac{\frac{\Lambda}{\sqrt{Z}} v_{\rm EW}}{m_h^2}\sim \frac{m_s}{m_h}, \eeq which becomes independent of $\Lambda$. Through the mixing $s$ couples to the Standard Model particles. Thus, it can be produced in accelerators or stars if the mixing is large enough, and if $s$ is in the relevant mass range. The mixing-induced coupling can also lead to a fifth force mediated by $s$. For instance the $s$ nucleon coupling is $\approx \sin \theta_{hs} g_{hNN}\approx 10^{-3} \sin \theta_{ hs}$~\cite{Piazza:2010ye}. 

Neglecting the model-dependent $\Psi$-mediated interaction for simplicity,\footnote{Taking into account the $\Psi$ interaction, $s$ can, for instance, decay into two photons via a loop, contributing to the decay rate as $\sim \frac{1}{\mathcal{O}(10^{5-6})} m_s^3 / f_a^2\propto Z^{-2} m_s$, which is of the same order as the contribution from Higgs mixing. It does not change the parameter region significantly.} the particle $s$ decays in the PQ broken phase through two main channels: into two axions via the self-coupling $\lambda$, and into lighter Standard Model particles via the Higgs portal coupling.
The former channel has a decay rate of 
\beq
\Gamma_{s \to aa} = \frac{m_s^3}{32 \pi f_a^2} \propto Z^{-2} m_s.
\eeq
The latter channel, for instance, decay into electron-positron pairs, has a decay rate of
\beq
\Gamma_{s \to e \bar e} = \frac{\theta_{hs}^2 y_e^2}{8\pi} \propto Z^{-1} y_e^2 m_s,
\eeq
with $y_e$ being the electron Yukawa coupling, 
and decay into photon pairs is given by
\beq
\Gamma_{s \to \gamma\gamma} \sim \frac{\theta_{hs}^2 m_s^3}{\mathcal{O}(10^{3-4}) \pi^5 v^2} \propto Z^{-2} m_s,
\eeq
where more precise discussions can be found in Ref.~\cite{Leutwyler:1989tn}. Interestingly, the decay rates satisfy $\Gamma_{s \to aa} \sim \Gamma_{s \to \gamma\gamma}$ for $\lambda_P \sim 1/Z$ and $\Lambda = 1\,\TEV$.
For $m_s < \MEV$ and thus $f_a \gg 10^9\GEV$, the lifetime of $s$ can exceed the age of the Universe, making the PQ Higgs a viable dark matter candidate.

Given these relations, the parameter region for $s$ is shown in Fig.~\ref{fig:1} on the $m_s$-$\theta_{hs}$ plane. 
The constraints include data from accelerator experiments (gray region at the top)~\cite{NA62:2021zjw,NA62:2020pwi,NA62:2020xlg,CHARM:1985anb,LHCb:2012juf,L3:1996ome}. The star cooling bound is recast from Ref.~\cite{Dev:2020eam} (blue region in the middle). The fifth-force bound is recast from Ref.~\cite{Salumbides:2013dua} (see also \cite{Adelberger:2009zz,Salumbides:2013aga}), neglecting the coupling to electrons. 
Assuming $s$ constitutes DM, its loop-induced decay into a pair of photons is also constrained~\cite{Yin:2024lla,Wadekar:2021qae,Janish:2023kvi,Calore:2022pks,Sun:2023acy,Caputo:2022mah,Nakayama:2022jza,Carenza:2023qxh}. The combined limits are taken from \cite{Sakurai:2021ipp}, \cite{Salumbides:2013dua}, and \cite{AxionLimits}. 
To account for uncertainties in the order-of-magnitude estimates, the prediction of the FIPQ model (and CP-even Higgs \cite{Sakurai:2021ipp,Haghighat:2022qyh}) is represented as a broad band spanning two orders of magnitude.

\paragraph{Quality Problem}

The axion solution to the strong CP problem is known to face the ``quality problem", raising the question of why the PQ symmetry must be extraordinarily precise to prevent CP violation in the QCD sector. Even if a global PQ symmetry is imposed, two sources of explicit PQ symmetry breaking remain: (1) gravitational effects and (2) non-gravitational effects.

For (1), global symmetries are generally expected to be broken by quantum gravity effects such as wormholes~\cite{Lee:1988ge,Giddings:1987cg,Abbott:1989jw,Kallosh:1995hi,Alvey:2020nyh}. Unfortunately, the model described here may not, by itself, alleviate the quality problem arising from (1), as introducing a large wave function renormalization constant does not alter the universal gravitational coupling. However, the harmful instanton size of a wormhole is of the order $\sim 1/M_{\rm pl}$~\cite{Kallosh:1995hi,Alvey:2020nyh}, indicating that this effect is UV-sensitive. Here, $M_{\rm pl} \approx 2.4 \times 10^{18}~\mathrm{GeV}$ is the reduced Planck mass. Consequently, I assume a UV completion of the model that resolves this issue, such as a large-volume extra dimension~\cite{Kallosh:1995hi} (see footnote.~\ref{ft1}) and a large non-minimal coupling~\cite{Hamaguchi:2021mmt} (although this may not work for FIPQ, e.g. \Sec{2-1})

For (2), an example is the CP-violating small QCD instanton contributions~\cite{Dine:1986bg,Kitano:2021fdl}, which can arise from higher-dimensional embeddings~\cite{Poppitz:2002ac,Gherghetta:2020keg,Kitano:2021fdl}, larger gauge theories~\cite{Agrawal:2017ksf,Agrawal:2017evu,Fuentes-Martin:2019bue,Csaki:2019vte,Takahashi:2021tff,Babu:2024udi}, or composite dynamics~\cite{Gherghetta:2020ofz,Aoki:2024usv}. Recently, it has also been suggested that large non-minimal couplings of any scalar field may enhance the small instanton effect due to the threshold effect~\cite{Wada:2024txn}.

The FIPQ hypothesis can alleviate the issue posed by (2). The PQ-breaking term generated at scales higher than $\Lambda$ by non-gravitational interactions is suppressed when transforming to the canonical basis, for instance:
\beq
{\cal L}_{\rm PQV} \supset \frac{(\Phi)^d}{M^{d-4}} + h.c. \to Z^{-\frac{d}{2}} \frac{(\Phi)^d}{M^{d-4}} + h.c. \sim \frac{\Lambda^d}{|M|^{d-4}} 2 \cos{\left(d \frac{a}{f_a} + \theta_{\rm CPV}\right)},
\eeq
where $d$ is the dimension of the term, and $\theta_{\rm CPV}$ is the CP-violating phase in the coupling. 

This suppression does not depend on the size of $f_a$. The contribution should be much smaller than the QCD contribution to axion, to have axion solving the strong CP problem:
\beq
\(\frac{\L}{M} \)^d \frac{M}{\chi_0}\lesssim 10^{-10}
\eeq
with $\chi_0\approx (0.08\GEV)^4$ being the topological susceptability. 
 For $d_{\rm FIPQ} \gtrsim 6$ and $M \sim M_{\rm pl}$, this can be satisfied for any $f_a$. For comparison in the ordinary PQ case, $\(\frac{f_a}{M}\)^d \frac{M^4}{\chi_0}\lesssim 10^{-10}$  is needed, which leads to $d_{\rm PQ}\gtrsim 14$ for $f_a=10^{12}\GEV$ and $M=M_{\rm pl}$.

This suppression is not an artifact of choosing the basis. It can also be understood by considering small QCD instanton effects in the canonical basis.
Given that small instantons are generated at high energy scales, contributions involving $n$ PQ Higgs fields must be accompanied by small couplings suppressed by $1/Z^{n/2}$. For a concrete example, the PQ quark masses contribute to the small instanton, and thus the small instanton is suppressed by the product of the small PQ quark masses $\sim \mathrm{TeV}$ (and by the small couplings for including $s$ loops).

\section{DM Cosmology in FIPQ }
\lac{2}
The cosmology of this scenario is also different from that of usual PQ models. In particular, one should take into account the dynamics of light $s$ in contrary to the conventional, especially the preinflationary PQ breaking, scenarios. 
Let us discuss DM production by considering inflationary cosmology. I will show that there are three kinds of viable scenarios for the DM cosmology.

\subsection{FIPQ before Big-Bang}
\lac{2-1}
Before discussing the DM cosmology, let us consider the period of inflation. To be concrete, I assume that the inflationary Hubble scale, $H_{\rm inf},$ is a free parameter\footnote{There are various inflation models that the dynamical mass scale is around or below the EW scale consistent with the FIPQ assumptions, e.g.,~\cite{Bezrukov:2007ep, Bezrukov:2008ej, Takahashi:2019qmh, Yin:2022fgo}, although we may not need to impose the assumptions in the inflation sector. The small radiative correction to the EW scale can be maintained if the inflaton-Higgs coupling is small, which predicts low reheating temperature, e.g.\,~\cite{Jaeckel:2020oet}.   }, while I will denote the Hubble parameter at an arbitrary cosmic time, $H$. 
The inflaton is not directly coupled to the PQ Higgs sector, and thus, the reheating caused by the inflaton decay in the inflaton matter-dominated epoch dominantly occurs in the Standard Model sector with a time-independent decay rate. I further assume that the inflation lasts long enough. I will not discuss the phenomena during preheating, such as the case that the coherent oscillation of $s$ starts during this short period, which is model-dependent. 

Usually, if the inflation scale is high, the PQ Higgs boson or any other scalars would acquire a sizable Hubble-induced mass via the non-minimal coupling, which is not forbidden by any symmetry. Thus 
\beq
{\cal L}\supset - \x |\F|^2 R \to \d V= 3 \x H_{\rm inf}^2 \ab{\F}^2
\eeq
during the inflation. In the FIPQ scenario, however, the non-minimal coupling $\x$ is naturally $\O(Z^{-1}) $ in the canonical basis.
This results in the non-minimal coupling and Hubble induced mass also being highly suppressed.

Since $H_{\rm inf}\gg \sqrt{|\x|} H_{\rm inf}$, the PQ symmetry rarely restores during the inflation in the observable Universe irrelevant to the sign of $\x$ and the size of $H_{\rm inf}$.  To see this, let us consider the case that the Hubble induced mass is larger than the vacuum mass $m_\F$, since the other case is trivial. 
In this case, the stochastic random jump of $\D \F\sim \frac{H_{\rm inf}}{(2\pi)}$ per a Hubble time is always larger than the effective mass $\sim \sqrt{|\x|} H_{\rm inf}\sim \sqrt{Z^{-1}} H_{\rm inf}$ in our scenario. If inflation lasts long enough, the random jump would balance with the classical motion, and the system forms an equilibrium distribution, with the typical field value satisfying $\vev{V[\F]}\sim \frac{3H_{\rm inf}^4}{8\pi^2}$~\cite{Starobinsky:1986fx,Starobinsky:1994bd,Hardwick:2017fjo, Graham:2018jyp,Guth:2018hsa,Ho:2019ayl,Alonso-Alvarez:2019ixv}, which leads to 
\beq
\laq{F}
\vev{|\F|^2}_{\rm inf}\sim  Z\sqrt{\frac{3H_{\rm inf}^4}{ 8\pi^2}}.
\eeq
Here I have set $y\sim 1/\sqrt{Z}, \l\sim 1/Z^2.$ Note that this distribution does not depend much on the sign of the Hubble induced mass as long as $|\x|\ll1$. 
Let us for simplicity consider the range $\sqrt{\vev{|\F|^2}_{\rm inf}}\ll M_{\rm pl}$, 
which leads to 
\beq\laq{amp}
H_{\rm inf} \ll 10^{13}\GEV \frac{\L}{1\TEV}\frac{10^{9}\GEV}{f_a}.
\eeq
This will be the parameter region of interest (c.f. the constraint from the non-observation of the tensor mode reads $H_{\rm inf}\lesssim 4.7 \times 10^{13}\GEV$~\cite{Planck:2018jri,BICEP:2021xfz}.).  

At the end of inflation, $\F$ has the typical field value \Eq{F} in the observable Universe, while in each Hubble patch, it has small fluctuation from the value by $\O(H_{\rm inf})$.\footnote{The contribution to isocurvature mode is suppressed due to the afterward dynamics, such as the fluctuation damping due to the large positive thermal mass squared in (Ib), the large oscillation amplitude induced by the negative thermal mass in (IIa), or small inflationary Hubble parameter in (IIb). In particular in (IIa) scenario, the anhormonic effect may enhance the isocurvature contribution~\cite{Kobayashi:2013nva}. Even if the anharmonic effect dominates the abundance estimation, the contribution to the power spectrum of the isocurvature mode should be $\sim \frac{H_{\rm inf}^2}{\(2\pi\)^2 \vev{|\F_{\rm inf}|^2}}\sim \frac{1}{(2\pi)^2 Z},$ which may be probed in the future for $f_a=10^{8-9}\GEV.$
}
Then, the reheating of the Universe occurs. During the reheating phase, we also have the Hubble induced mass to $\F$, but this mass is again much smaller than the Hubble parameter and we neglect the impact in the following discussion.  
There are two scenarios: \begin{description} \item (I) The reheating phase is short enough that every relevant dynamic happens during the radiation-dominated Universe. 
\item (II) The reheating phase lasts very long, affect the relevant dynamics.
\end{description}

\subsection{FIPQ in Thermal Environment}
\lac{2-2}
I further review some thermal behavior of the PQ field in preparation for discussing DM cosmology. 

In the thermal environment characterized by the cosmic temperature $T$, the thermal effect modifies the PQ Higgs potential as
\beq
V_T = \frac{T^4}{2\pi^2} \sum_i g_i J_i\left(\frac{m_i^2}{T^2}\right),
\eeq
where
\beq
J_{i}(z^2) = (-1)^{2s_i} \int_0^{\infty} dx\, x^2 \log\left(1 - (-1)^{2s_i} e^{-\sqrt{x^2 + z^2}}\right),
\eeq
with $s_i$ being the spin of particle $i$ and $g_i$ being the degrees of freedom. In the case of $i = \Psi$, a color triplet fermion, we have $g_i = 12$, $s_i = 1/2$, and $m_\Psi = y \Phi$. Note that PQ fermions and the Standard Model-like Higgs boson have much stronger Standard Model interactions, which contribute to the dominant thermal mass $\sim \mathcal{O}(T)$ to $m^2_i$. In other words, we perform a daisy resummation to ensure $z^2 > 0$ in $J_{i}(z^2)$ even with a negative $\lambda_P$, with $m_i^2 \to m_i^2 + c_i T^2$, where $c_i T^2$ is the thermal mass, $c_i = \mathcal{O}(1)$. With this resummation, the square root in the exponent is real.

When $\Phi$ is near the origin, by expanding $z^2$ in $J_i$, the thermal mass term is obtained  
\beq
V_T \sim \left(\frac{g_\Psi y^2}{48} + \frac{\lambda_P}{6}\right) T^2 |\Phi|^2 =\O(Z^{-1})T^2 |\F|^2.
\eeq
When $T$ is sufficiently large, the PQ Higgs acquires a large thermal mass squared with two possible signs:
\begin{description}
    \item[(a)] Positive thermal mass squared: $\lambda_{P} > -c \frac{g_\Psi}{8} y^2$
    \item[(b)] Negative thermal mass squared: $\lambda_{P} < -c \frac{g_\Psi}{8} y^2$
\end{description}
Here, $c$ is introduced as a factor, which is not exactly unity due to the daisy resumption and higher order term contribution. For $g_\Psi = 12$, we have $c \approx 1.6$ for $c_i = 1$. In the following, I take $c = 1$ and use the approximate formula for studying the dynamics.\footnote{Strictly speaking, with the daisy resummation, the high-mass expansion rather than the low-mass expansion is accurate. I checked that
\[
J_B(z^2) = -\sum_{n=1}^{n_{\rm max}} \frac{1}{n^2} z^2 K_2(n \sqrt{z^2})
\]
and
\[
J_F(z^2) = -\sum_{n=1}^{n_{\rm max}} \frac{(-1)^n}{n^2} z^2 K_2(n\sqrt{z^2})
\]
approximate the potential very well with $n_{\max} \sim 10$. Using this form, one can perform a more precise numerical simulation. For simplicity, however, I use the simple quadratic expansion in this paper.}

\subsection{Case (Ia) does not work for DM production}
\lac{2-3}
 In this subsection, I consider the parameter region $(a)$ with the positive thermal mass squared and the scenario (I) that reheating ends early enough.  

Although $s$ has nonvanishing value from the inflationary dynamics, it settles in the potential minimum in the early Universe. 
This happens if the thermal mass is larger than the Hubble parameter
\beq\laq{onset}
  \frac{1}{\sqrt{Z}} T \gtrsim H.
\eeq
 Again, I used $y\sim 1/\sqrt{Z}, \l\sim 1/Z^2.$
Then, the symmetric phase of the PQ is obtained since the expectation value of $\F$ is zero.

Later, the Universe cools down, and the vacuum mass squared of $\F$ becomes important, and the PQ symmetry breaks again. 
By comparing the thermal mass and the vacuum mass $m_s$, we find that the phase transition occurs at \beq T = T_{\rm PQB} \sim \Lambda. \eeq Afterwards, the PQ fermions, which were massless, acquire masses of order $\L\sim \TEV$, which decouples. 

Although the thermal mass estimation is not directly relevant to whether $s$ or $a$ is thermalized, the thermalization rate is dominantly determined by the interaction with the massless PQ fermions, e.g., $\Psi \bar\Psi \to s g$, where $g$ is a gluon. This rate is given by
\beq
\Gamma_{\rm th} \sim \frac{3y^2}{32\pi^3} T \sim \frac{3\Lambda^2}{32\pi^3 f_a^2} T,
\eeq
which is a contribution of $\mathcal{O}(Z^{-1})$ and is much faster than the contribution via the Higgs portal coupling, which is $\propto \lambda_P^2 T \propto 1/Z^2$. The fermion decouples from the system when $T < \Lambda$, the vacuum mass. Thus, we compare the thermalization rate with the Hubble parameter at $T \sim \Lambda = \mathrm{TeV}$. From this, $s$ is not thermalized if
\beq
\laq{fath}
f_a \gtrsim 10^9\GEV \sqrt{\frac{\mathrm{TeV}}{\Lambda}}.
\eeq
Even if $s$ is not thermalized, the number density produced from the thermal scattering can be estimated as 
$
 n_s\sim \left.\frac{\G_{\rm th}}{H} T^3\right|_{T\sim \L}.
$
This gives
\beq \frac{n_s}{s}\sim 0.001\(\frac{10^{10}\GEV}{f_a}\)^{2}.\eeq
In addition, there is a contribution from the coherent oscillation from the survival of the remnant of the stimulated emission~\cite{Nakayama:2021avl}. 
The subsequent decay of $s$ contributes too much for consistent BBN and CMB data: 
e.g., $n_s m_s/s \ll 10^{-14}\,\GEV$ for a non-relativistic $s$ (see Refs.~\cite{Kawasaki:2017bqm,Poulin:2016anj}) at the BBN epoch. 
For $f_a\sim 10^{10}\GEV$, this is problematic. Either $f_a\gg 10^{10}\GEV$ which implies $s$ rarely decays, 
or $f_a \ll 10^{10}$, where $s$ is shorter lived than BBN, remains. 
In the fomer case with $f_a\gg 10^{10}\GEV$, there is an overproduction of the axion DM from cosmic string network (see \Eq{string}). In the latter case the axion DM is subdominant and the DM is not enough.

This implies that (Ia), i.e., the symmetry breaking that happens in the radiation-dominated Universe, does not work for the DM production of either $s$ or $a$ in the current framework.\footnote{The inflaton stimulated decay to produce $a$ DM in $f_a\sim 10^{8-9}\GEV$ could work in (Ia) scenario~\cite{Moroi:2020has, Moroi:2020bkq, Choi:2023jxw}, by relaxing the assumption that the inflaton does not directly couple to the PQ sector. Stimulated decay or scattering can also work for producing $\O(1-100)\EV$ mass range PQ Higgs DM~\cite{Yin:2023jjj,Sakurai:2024apm} by assuming certain couplings.}

\subsection{Case (IIa): slim axion DM from fat string network}

\lac{2-4}
If $\U(1)$ symmetry is spontaneously broken, we have cosmic strings. 
Those strings form a scaling solution and evolve such that $\O(1)$ string typically exists in a single Hubble patch. Such strings radiate axions and can make axions become dominant DM depending on the decay constant. 
From Fig.\ref{fig:1}, despite the severe bound from star cooling, there is a narrow window around $m_s \sim 10\,\KEV$, where $f_a \sim 10^{10}\,\GEV$. Interestingly, this coincides with the parameter region for the axion DM from the string network~\cite{Saikawa:2024bta,Kim:2024wku,Buschmann:2024bfj,Kim:2024dtq}, following which we expect 
\beq 
\laq{string}
m_a \sim \O(10)\mu\EV \(\frac{\Omega_a}{0.12 h^{-2}}\)^{-1/1.17} \(\frac{\log{(m_s/H_{\rm QCD})}}{70}\)^{1/1.17},
\eeq 
if the string contribution is dominant. Here, $H_{\rm QCD}$ is the Hubble parameter when the mass of the axion becomes comparable to $H$, and the conventional logarithmic dependence is $\log(m^{\rm PQ}_s/H_{\rm QCD}) \sim 70$.
For instance, in \cite{Saikawa:2024bta}, the value for the sample parameter is $95{-}400\m\EV$, indicating $f_a \sim 10^{10-11}\GEV$, where the theoretical uncertainties include explorations from much smaller $\log(m_s/H_{\rm QCD})$ in numerical simulations.

Indeed, the scaling solution of the string holds in the matter-dominated Universe. Therefore, in the case (IIa), where we have late-time reheating, we can simply dilute the thermally/non-thermally produced $s$ particles around the dilute plasma temperature $T\sim\L$ to have a consistent cosmology with the BBN and CMB. 
Let the reheating temperature be $T_R$, satisfying $T_R \ll T_{\rm PQB} \sim \L$. Then the thermally produced $s$ particles until $T \sim \L$ are diluted by a factor of $\(\frac{T_R}{\L}\)^7$. Thus, for $T_R \lesssim 0.1\L$, we have a sufficiently suppressed $s$ abundance to save the cosmological problem caused by the late time decay of $s$. The axion DM is then produced from the string network as well. 

In my setup, $m_s$ is much smaller than the conventional value
\beq
\laq{predfat}
\boxed{\log{\(\frac{m_s^{\rm FIPQ}}{H_{\rm QCD}}\)} \sim 30\to m_{a}^{\rm FIPQ}\sim \frac{1}{2} m_{a}^{\rm PQ}.}
\eeq
This still largely overlaps with the conventional prediction, $95{-}400\m\EV$~\cite{Saikawa:2024bta}, but we have a slightly smaller preference for the axion mass. For instance, QUAX~\cite{QUAX:2024fut}, MADMAX~\cite{Beurthey:2020yuq}, ALPHA~\cite{Lawson:2019brd}, and ADMX~\cite{Stern:2016bbw} can distinguish the scenario if the mass is in the range of $40{-}95\m\EV$.  
A better understanding of the axion mass prediction from the string network would strengthen distinguishability as long as $m_s$ affects the axion abundance. Indeed, if I refer to the more recent study \cite{Buschmann:2024bfj}, which predicts $m_{a}^{\rm PQ} = 45{-}65\m\EV$, the FIPQ scenario is already distinguishable since it predicts $m_a^{\rm FIPQ} \approx 20{-}30\m\EV$. However, if the domain wall contribution is significant, as preliminarily discussed in \cite{Buschmann:2024bfj}, the prediction of \Eq{predfat} may change depending on the sensitivity of the axion abundance to $m_s$.

\subsection{Case (Ib): Axion DM from PQ Higgs fragmentation}
\lac{2-5}
In the parameter region of $(b)$ and scenario (I), after reheating, the $\F$ field slow-rolls toward the direction of the phase, which is stochastically chosen during inflation. The onset of oscillation of the Higgs occurs at the temperature 
\beq
T=T_{\rm osc} \sim \frac{M_{\rm pl}}{\sqrt{Z}}.
\eeq
At this moment, the thermal potential has a minimum at 
\beq
|\F|^{\rm min}_{T_{\rm osc}} = \O(0.1)M_{\rm pl},
\eeq
which is independent of $Z$.\footnote{The PQ fermions, as well as the Standard Model-like Higgs field, obtain a mass squared contribution from the expectation value, $\O(0.01)Z^{-1} M^2_{\rm pl}$, which is smaller than or comparable to $T^2_{\rm osc}$. Therefore, the previous analysis of the thermal potential remains valid.} Here and in the following order estimation, I approximate the negative thermal mass parameter squared as $\sim -\frac{\lambda_P}{6} T^2 \sim -\frac{T^2}{6 Z}$. Then the size of the initial field value squared $ \sim \vev{|\F|^2}_{\rm inf} \ll M_{\rm pl}^2$, due to \Eq{amp}, can be neglected in the following discussion of the field dynamics. Satisfying \Eq{amp}, the fermion and Standard Model-like Higgs $\F$ induced masses below $T$, making the thermal potential estimation consistent.\footnote{Another consistency condition for having a radiation-dominated Universe at this period reads\footnote{If the reheating temperature does not satisfy this condition, the $s$ abundance is suppressed due to further entropy production.} $T_R \gtrsim 10^{12}\GEV \frac{10^{9}\GEV}{f_a}\frac{\L}{1\TEV},$ which implies \beq H_{\rm inf} \geq \sqrt{\frac{g_{\star} \pi^2}{90 M_{\rm pl}^2}} T^2_R \gtrsim 10^6\GEV \(\frac{10^{9}\GEV}{f_a}\frac{\L}{1\TEV}\)^2. \eeq There is a large parameter region satisfying this condition.}

Neglecting momentum-dependent fluctuations, $|\F|_{T}^{\rm min}$ is always comparable to the oscillation amplitude because both scale with $R^{-1}$, where $R$ is the scale factor, assuming the adiabatic invariant is conserved.\footnote{Similar dynamics were noted in Ref.~\cite{Yin:2024trc} to reduce neutrino-right-handed-neutrino mixing by coupling the scalar to the right-handed neutrino making it have a time-dependent mass. In that case, the oscillation originates from a large field value with the thermal potential, and no lighter bosonic particle exists. Fragmentation induced by tachyonic instability and stimulated emission does not occur.} This implies that $s$ can return close to the potential origin even with the redshift. Given that the adiabatic condition is slightly violated, $s$ overshoots the origin.  

\begin{figure}[!t] 
\begin{center}
\includegraphics[width=75mm]{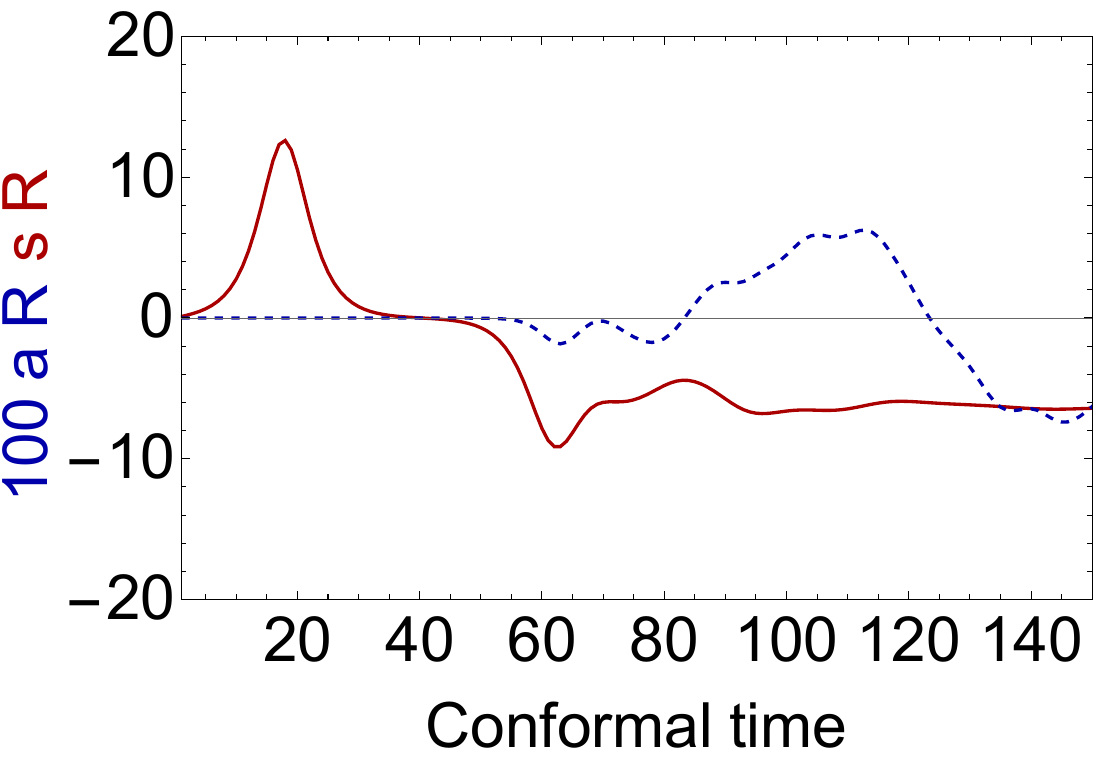}
\hspace{3mm}
\includegraphics[width=80mm]{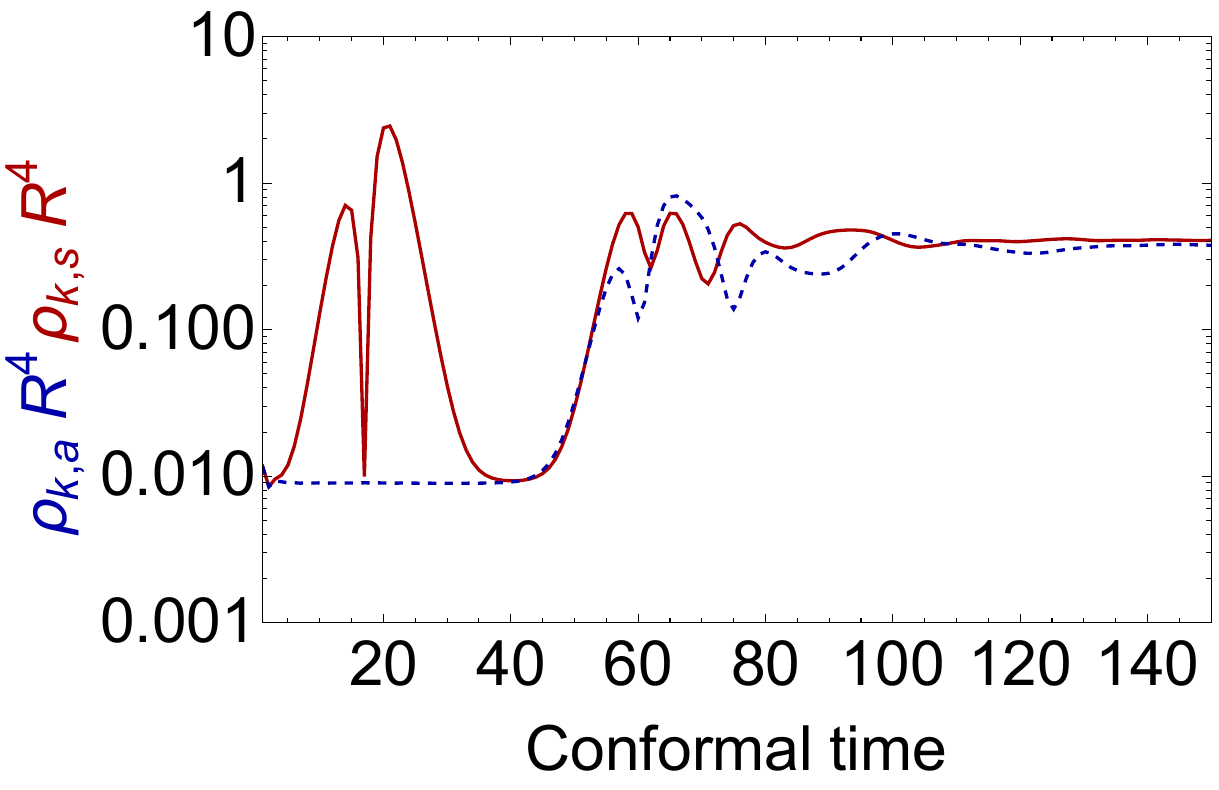}
\end{center} 
\caption{Lattice simulation for the spatially averaged $\Re \F \times R/\sqrt{2}$ (red solid line) and $100 \times \Im \F \times R/\sqrt{2}$ (blue dashed line) evolutions [left panel]. The overshoot occurs only once. Around the overshoot, $a$ particles (and $s$ particles) are significantly produced, as shown by the kinetic energies $\times R^4$ denoted by the same line styles [right panel]. The initial condition for the homogenous mode is set such that the average of $\Re \F$ is slightly displaced from the origin, while $\Im \F=0$.
A machine unit is used for the dimensionful parameters.}
\label{fig:lat} 
\end{figure}

\begin{figure}[!t] 
\begin{center}
\includegraphics[width=75mm]{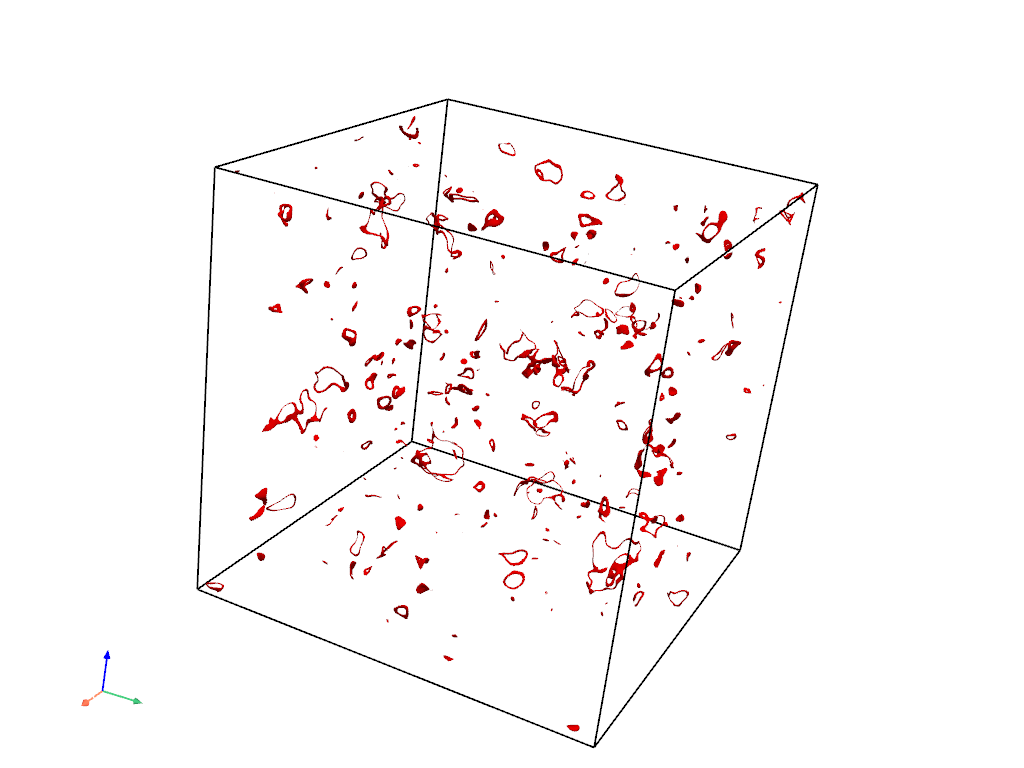}
\hspace{3mm}
\includegraphics[width=75mm]{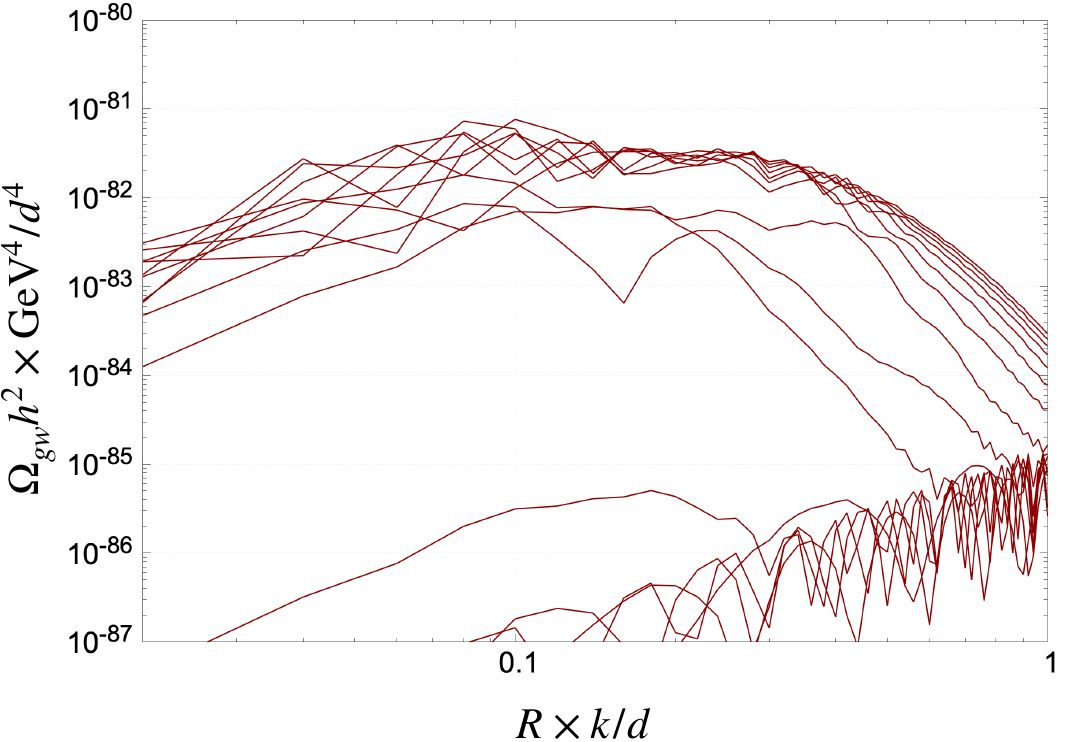}
\end{center} 
\caption{The snapshot at conformal time = $141/d$ [left panel], and the spectra of the gravitational wave at different conformal time slices [right panel], where the horizontal axis denotes the comoving momentum.  Again, $d$ is the machine unit. The simulation correspnds to (Ib), where there is no PQ phase transition. 
 }
\label{fig:lat2} 
\end{figure}

In addition, the potential origin has a negative curvature, and various non-linear and non-perturbative effects, such as tachyonic instability~\cite{Felder:2000hj,Felder:2001kt}, become important. To address these effects, I perform a classical $512^3$  lattice simulation (more detailed studies will be presented in~\cite{WY}) using {\tt Cosmolattice}~\cite{Figueroa:2020rrl,Figueroa:2021yhd}, incorporating the scale factor-dependent negative mass squared $\propto R^{-2}$ in a two-scalar-field system for numerical simulation. Some results are shown in Figs.\ref{fig:lat} and \ref{fig:lat2}.\footnote{The setup uses $\Re \F = 0.1 d$, $\Im \F = 0 d$ as the initial condition, $H = d R^{-2}$, $\lambda = 0.001$, $k_{\rm IR}=0.02 d/R$, which corresponds to the box size, and a time-dependent negative mass squared $m_\F^2 = -0.08d^2 R^{-2}$. The conformal time $\times d= R$ in the simulation. 
The initial fluctuation has a reduced power spectrum of $0.01k^3 /(4\pi^2)$ and cut off at the momentum of $d$, which is the gaussian fluctuation. For inflationary fluctuation~\cite{Gonzalez:2022mcx,Kitajima:2023kzu}, the conclusions do not change as will be shown in Fig.~\cite{WY}. 
We use the Minkowski fluctuations provided by the code with a positive mass squared set by hand~\cite{Felder:2000hj,Felder:2001kt} (see also the discussions in \cite{Kitajima:2022jzz, Gonzalez:2022mcx,Kitajima:2023kzu}). }  

I numerically find that with an $\O(1)$ oscillation, $s$ settles into a minimum, and the condensate fragments into axions and PQ Higgs particles with comparable energy densities. 
This phenomenon may be understood as follows. 
In the Boltzmann picture of axion production, stimulated emission is important if $n^{(0)}_s/(Hf^2_{a}(T)) \gg 1$~\cite{Moroi:2020has}, where $n^{(0)}_s$ is the condensate density, $f_a(T)$ ($m_s(T)$ appearing later) is the temperature-corrected decay constant  (mass). Here, the produced axions carry momentum $p_a \sim m_s(T)/2$, which redshifts as $p_a \propto R^{-1}$ after production.  
If the condensate remains, $f_a(T) \propto R^{-1}$, $n^{(0)}_s \propto R^{-3}$, $m_s(T) \propto R^{-1}$, and $H \propto R^{-2}$. Since $m_s(T)\sim H \AND n^{(0)}_s\sim m_s(T) f_a(T)^2,$ $n^{(0)}_s/(Hf^2_{a}(T)) \sim 1$ at the onset of oscillation, stimulated emission remains important afterward. The mass of $s$ and the momentum of produced axions redshift similarly, leading to the equilibrium gradually.  

After equilibrium, the energy densities redshift as radiation. Introducing a tiny constant mass term that becomes relevant later in the simulation shows no further significant axion production.  

Interestingly, using equilibrium and the fact that total energy $\times R^4$ is conserved, I find 
\beq
\rho_s \sim \rho_a \sim 10^{-2}T^4,
\eeq
which is not too small compared to the thermal plasma. Identifying the thermal mass as 
$\sqrt{-2 \(\frac{g_\Psi y^2}{48} + \frac{\lambda_P}{6}\right) T^2}$ $\propto Z^{-1/2} T$, I obtain  
\beq
\frac{n_{a}}{s} \sim \frac{n_{s}}{s} = \O(0.1)\frac{\sqrt{\l} (s^{\rm min}_{T_{\rm osc}})^3}{2 g_{\star s}T_{\rm osc}^3 \pi^2/45} 
= \O(10^{-4}) Z^{1/2}.
\eeq
This leads to the abundance
\beq
\laq{sab}
\boxed{\Omega_{s} = m_{s} \frac{n_{\F}}{s} \frac{s_0}{\rho_c} \sim 10^{5} \L,~~~~\Omega_{a} = m_{a} \frac{n_{a}}{s} \frac{s_0}{\rho_c} \sim 10^5 \frac{\chi_0^{1/2}}{\L}.}
\eeq
Interestingly, these results do not depend on $Z$. Furthermore, 
\beq
\Omega_{a} \sim 1 \({\frac{\chi_0}{(0.08\GEV)^4}}\)^{1/2}\frac{1\TEV}{\L},
\eeq
which is close to the desired value~\cite{Planck:2018vyg}.

Again, we need to deal with the overproduction of $s$. One simple way to avoid the cosmological problem is to consider 
$
f_a \lesssim 10^{9}\GEV,
$
such that $s$ is thermalized due to the fermionic interaction before the fermion decoupling (see \Eq{fath}). 
Under this condition, $s$ is typically heavier than the electron and can naturally decay before the BBN, either without or with moderate entropy dilution. This leads to the prediction that 
\beq
m_a \gtrsim 6\,\mathrm{meV}.
\eeq
Interestingly, $s$ in this region may be probed in future accelerator experiments, for example, via meson decays.

\paragraph{Small string loops, strong gravitational waves, and miniclusters}
Although the axion has a very suppressed decay constant in the vacuum in the scenario (Ib), the thermally induced negative and large mass squared causes $s$ to begin to oscillate around the potential minimum at $|\F|_{T_{\rm osc}}^{\rm min}=\O(0.1)M_{\rm pl}$. At this early epoch the expectation value $\vev{\F}=\O(0.1)M_{\rm pl}$ and the axion acquires a very large decay constant around the Planck scale. 

According to the numerical simulation, despite the fact that the PQ symmetry remains broken, string loops are formed during the first few oscillations of $s$ (see the left panel of Fig.\ref{fig:lat2}.). This may be because $s$ overshoots the origin by $\O(1)$ times, during which fluctuations enhanced by tachyonic instability develop in different directions in the field space, leading to the formation of string configurations.  

These string loops have a long life time, similar to the melting topological defects scenarios~\cite{Babichev:2021uvl, Babichev:2023pbf, Dankovsky:2024ipq}.\footnote{The difference from the original melting cosmic string scenario is that we do not have a long cosmic string at the later time, but only long-lived small loops produced in the first few oscillations. 
As I will discuss in the last section, this does not introduce the domain wall problem even in DFSZ realizations. } 
Indeed, in the left panel of Fig.\ref{fig:lat2}, we have the snapshot at the conformal time $141$, which implies that $1/H\sim 1/2 $ box length, but we have much more subhorizon string loops.

Since the cosmic string of size of the inverse of the mass scale, which is the Hubble scale, is formed with the large tension, 
$
\m \sim \(\O(0.1) M_{\rm pl}\)^2,
$
which redshift afterwards, 
the gravitational waves are mostly produced at the formation.
For the gravitational amplitude, the prediction of the melting cosmic string~\cite{Emond:2021vts} should be adopted, \beq
\Omega_{\rm gw}\sim 10^{-5} \(\frac{G \m }{10^{-2}}\)^2,
\eeq
which does not depend much on the parameter choice. This agrees well with my simulation result in the right panel of Fig.\ref{fig:lat2}, where I used the machine unit $d$.  By taking $\mu = 0.08 /(\l 70^{2})/d^2\sim 0.02/d^2 $(because I checked that $R\sim 70$ leads to the dominant GW production), 
the formula gives $\Omega_{\rm gw}\sim 10^{-81}\frac{d^{4}}{\GEV^{4}}$.
The frequency depends on the parameter choice, which is \beq
f \sim \left.\frac{s_0^{1/3}}{s^{1/3}}H\right|_{T=T_{\rm osc}} \sim 10^4{\rm Hz} \frac{\L}{1\TEV} \frac{10^9\GEV}{f_a}. 
\eeq
For $f_a\sim 10^9\GEV$, given the large peak amplitude, the lower frequency tail may be probed by the gravitational wave detectors such as CE and ET.

\subsection{Case (IIb): PQ Higgs DM, axion DM, and co-DM}
\lac{2-6}
When the mass of the PQ Higgs is smaller than the electron mass, especially for $m_s \ll \MEV$, the lifetime can be much longer than the age of the Universe, and the PQ Higgs boson becomes a good DM candidate. Indeed, in the scenario (IIa), I found that the abundance of $s$ is irrelevant to the decay constant, \Eq{sab}, which is always too large under the assumption of a radiation-dominated Universe.
This led me to explore a matter-dominated Universe during reheating, which would dilute the abundance and may allow $s$ to become the dominant DM. In this case, however, the initial oscillation amplitude does not need to be as large as $\O(0.1)M_{\rm pl}$ since $H_{\rm osc}\gg T_{\rm osc}^2/M_{\rm pl}$, \Eq{onset} and $|\F|^{\rm min}_{T_{\rm osc}}\sim Z^{1/2} T_{\rm osc}$. I impose 
\beq
\laq{sosclarge}
(|\F|^{\rm min}_{T_{\rm osc}})^2 \gtrsim \vev{|\F|^2}_{\rm inf},
\eeq
so that the inflationary stochastic distribution of $\F$ can again be neglected at the onset of oscillation. If $(|\F|^{\rm min}_{T_{\rm osc}})^2 \ll \vev{|\F|^2}_{\rm inf}$, I numerically checked that the large-amplitude oscillation produces both $\Re \F$ and $\Im \F$ particles with the quartic potential due to the parametric resonance as pointed out e.g.~\cite{Tkachev:1995md, Greene:1997fu} (see also \cite{Daido:2017wwb, Daido:2017tbr, Co:2017mop,Harigaya:2019qnl,Moroi:2020has,Moroi:2020bkq,Nakayama:2021avl} for the axion/ALP DM production in the context of the parametric resonance). Subsequently, the system approaches the symmetric phase of PQ. 
In such a case, axion DM from the cosmic string network appears when the PQ phase transition occurs, yielding the same prediction as (Ib) with $f_a \sim 10^{10-11}\GEV$. 

In this section, I primarily focus on $f_a \gtrsim 10^{11}\GEV$, and thus concentrate on satisfying \Eq{sosclarge} to avoid axion overproduction. When I refer ``(IIb)", it indicates the region with $f_a \gtrsim 10^{11}\GEV$. 
Due to the large decay constant, thermal production is suppressed and will not be discussed. I will use the subscript ``$R$" to denote quantities at the end of reheating and ``osc" to denote quantities at the onset of oscillation of $s$. 

The condition \eq{sosclarge}, together with $f_{a} \gtrsim 10^{11}\GEV$, should be satisfied in two regimes:
\begin{description}
\item[(i)] $|\F|^{\rm min}_{T_{\rm osc}} \gg f_a,$ which leads to $T_{\rm osc} \gg \L$,
\item[(ii)] $|\F|^{\rm min}_{T_{\rm osc}} \sim f_a,$ which implies $T_{\rm osc} \lesssim \L$.
\end{description}
This categorization is based on whether the thermal mass squared $\sim Z^{-1}T_{\rm osc}^2$ is larger than the vacuum mass squared $\sim \L^4 / f_a^2 \sim Z^{-1} \L^2$. In the former case, the thermal misalignment~\cite{Batell:2021ofv,Batell:2022qvr} occurs during reheating phase. In the latter case, vacuum misalignment~\cite{Preskill:1982cy,Abbott:1982af,Dine:1982ah} with stochastically favored initial misalignment angle~\cite{Graham:2018jyp, Guth:2018hsa, Ho:2019ayl,Alonso-Alvarez:2019ixv} dominates.

Let us first consider the former regime (i). During reheating, the dilute plasma temperature scales as $T \propto R^{-3/8}$.  
At the end of reheating, $T = T_R$, when the energy density of the inflaton equals that of the radiation, $\rho_{R} = \frac{g_\star \pi^2 }{30} T_R^4$. During the reheating phase, the energy density evolves as $\rho_R (T/T_R)^{8}$. 
Let $H_R \approx \sqrt{\rho_R / 3M_{\rm pl}^2}$ be the Hubble parameter at the end of reheating. The Hubble parameter at a dilute plasma temperature $T$ is then given by $H = H_R (T/T_R)^{4}$. When the equality in \Eq{onset} is satisfied, $s$ starts to oscillate around the temperature-dependent minimum. Since the temperature decreases very slowly, the oscillation is almost coherent with a constant mass. In this case, significant axion production, as in scenario (Ib), does not occur, and the number of axions produced in this way is highly suppressed compared to the radiation-dominated scenario, as verified numerically. Given that the number density at the onset oscillation is $Z^{1/2}T_{\rm osc}^4$ and later it redshift as $R^{-3}$, I get 
\beq
n_s(T) \sim Z^{1/2} T_{\rm osc}^3\(\frac{T}{T_{\rm osc}}\)^{8}\sim \frac{f_a}{\L} T_{\rm osc}^3\(\frac{T}{T_{\rm osc}}\)^{8}.
\eeq

The DM abundance can then be obtained using 
\beq
\Omega_s = \frac{s_0}{\rho_0} m_s \frac{n_s[T_R]}{s[T_R]}:
\eeq
\beq
\Omega_s \sim 1 \(\frac{f_a}{10^{11}\GEV} \frac{T_R}{10^{4}\GEV}\)^{5/3} \(\frac{10^3\GEV}{\L}\)^{2/3}.
\eeq
The ratio of the scale factors is given by
\beq
\frac{R_{R}}{R_{\rm osc}} \sim 10^5 \(\frac{10^{11}\GEV}{f_a} \frac{10^4\GEV}{T_R} \frac{\L}{10^3\GEV}\)^{8/9},
\eeq
and the oscillation temperature is
\beq
T_{\rm osc} \sim 10^6\GEV \(\frac{10^{11}\GEV}{f_a} \frac{\L}{10^3\GEV}\)^{1/3} \(\frac{T_R}{10^4\GEV}\)^{2/3}.
\eeq

\begin{figure}[!t] 
\begin{center}
\includegraphics[width=80mm]{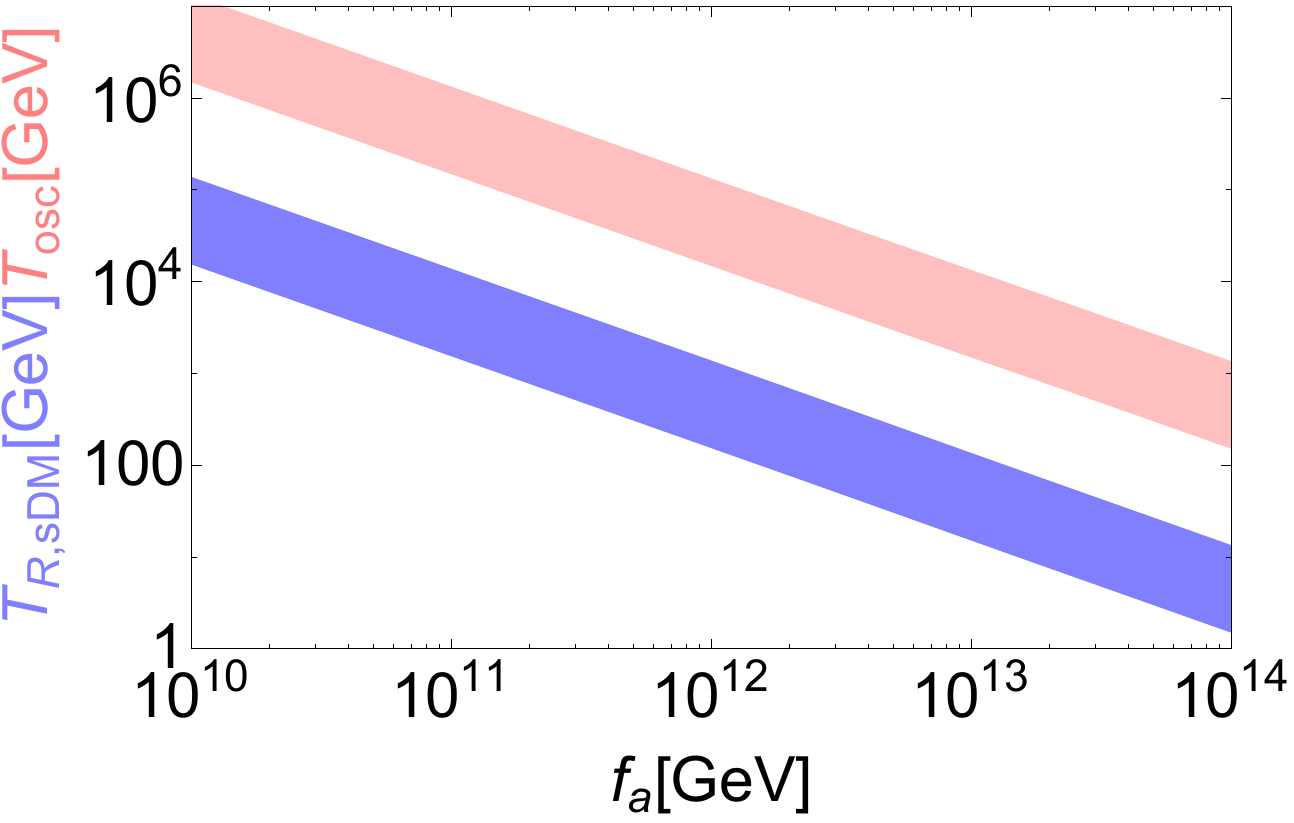}
\hspace{3mm}
\includegraphics[width=80mm]{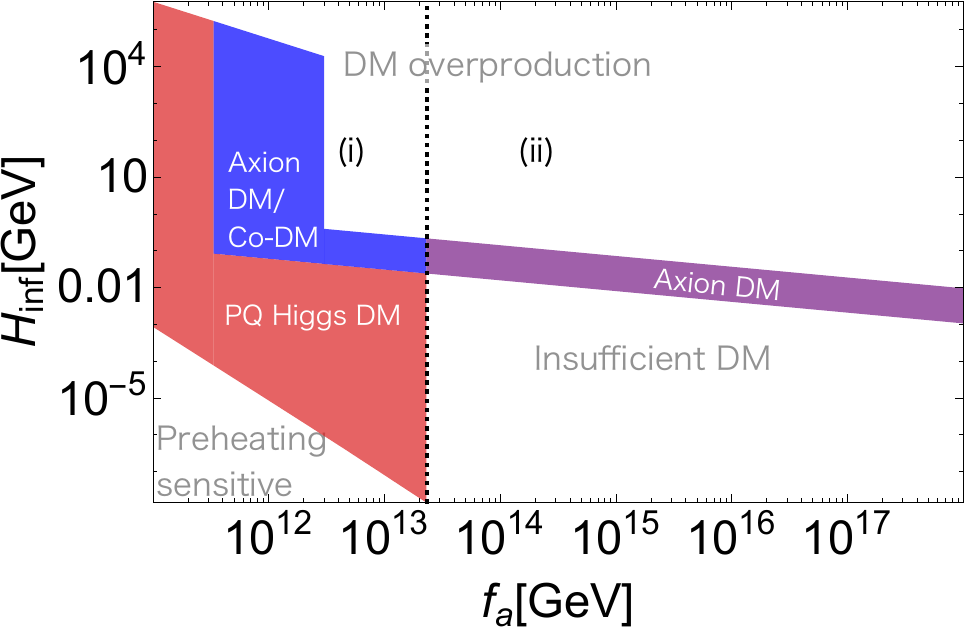}
\end{center} 
\caption{The reheating temperature prediction assuming the dominant thermal mass at the onset of oscillation, varying $f_a$ [left panel]. The temperature at the onset of oscillation is also shown. The parameter region in $f_a-H_{\rm inf}$ plane of scenario (IIb) [right panel]. The dominant DM candidate is indicated. The vertical dotted line separates the regime (i) and (ii).
In both figures, I fix $\L = 10^3\GEV$ and allow a factor of ten ambiguities for the vertical axis.}
\label{fig:2} 
\end{figure}

By fixing $\Omega_s = 0.3$, I obtain the prediction of $T_{R,\rm sDM}$, the reheating temperature required to explain the $s$ DM abundance for a given $f_a$, as shown in the left panel of Fig.~\ref{fig:2}. I also show the predicted $T_{\rm osc,sDM}$. For $f_a < 10^{13-14}\GEV$, we find regime (i), where thermal misalignment produces $s$ dark matter. The parameter region for (i) is shown on the region left to the vertical dotted line in the right panel of Fig.~\ref{fig:2}. 

In this scenario, $s$ dark matter production is successful and could be probed in future observations of X-rays, optical, and infrared light in galaxies, depending on the mass. The upper limit, $f_a \lesssim 10^{12}\GEV$, is determined by the condition \Eq{sosclarge} for PQ symmetry non-restoration. The lower limit ensures that $H_{\rm inf}$ is smaller than the thermal mass at the onset of oscillation. DM production in this regime is sensitive to the model-dependent preheating history.

For $f_a \gtrsim 10^{12}\GEV$, care must be taken to avoid overproduction of the QCD axion via vacuum misalignment.
In the case of $H_{\rm inf} \gg \Lambda_{\rm QCD}$, the axion is overproduced if $f_a \gtrsim 10^{12}\GEV$ and the initial misalignment angle is $\O(1)$. However, if $H_{\rm inf} \lesssim \L_{\rm QCD}$, this overproduction can be easily avoided due to the narrow stochastic distribution peaked around the CP-conserving minimum of the axion during inflation~\cite{Graham:2018jyp, Guth:2018hsa}. Interestingly, in some regimes, both $s$ and $a$ can be dominant DM. Detecting both DM candidates with the mass relation \Eq{masss} would be a smoking-gun signal (see \cite{AxionLimits} for axion DM reaches). 

In regime (ii), corresponding to $f_a \gtrsim 10^{13}\GEV$, we need to solve the axion overproduction problem. In my scenario, the simple way is to consider $H_{\rm inf} \lesssim \L_{\rm QCD}$, shown in the right-hand side of the right panel in Fig.~\ref{fig:2}.\footnote{Alternatively we can consider the reheating temperature to be lower than the QCD scale to dilute the axion abundance.}
Then $s$ vacuum misalignment is automatically highly suppressed~\cite{Ho:2019ayl}. Therefore, the QCD axion becomes the dominant DM, with $H_{\rm inf} \sim \L_{\rm QCD}$. Although $s$ is not the dominant DM, it could be probed through fifth-force experiments (see Fig.~\ref{fig:1}). Together with axion DM detection, observing a fifth force with the corresponding Compton length and sensitivity would also serve as a smoking-gun signal.

\section{Conclusions and discussion}
\lac{cd}

The QCD axion remains a compelling candidate for dark matter (DM) and a natural solution to the strong CP problem in the Standard Model. However, conventional models often require a high Peccei-Quinn (PQ) symmetry-breaking mass scale far above the electroweak (EW) scale to avoid constraints from star cooling, introducing potential challenges such as the PQ quality and hierarchy problems. 

In this work, I proposed an alternative framework where the PQ scalar field achieves a large decay constant through a large wave function renormalization constant, analogous to a feebly coupled gauge theory. Namely, the large decay constant is not a consequence of the large mass scale but the feeble interaction. 
This approach allows all other dimensionless parameters to remain $\mathcal{O}(1)$ and dimensionful parameters to stay close to the EW scale in the PQ sector before the kinetic normalization, and mitigate the PQ quality and hierarchy problems. 

The model predicts a light PQ Higgs boson with a mass $\sim (\mathrm{EW~scale})^2 / f_a$, where $f_a$ is the axion decay constant, and exotic particles inducing PQ anomaly with masses around the EW scale.

This framework introduces rich cosmologies:
\begin{itemize}
    \item Slim axion DM from a fat string network, with a factor of $\sim 2$ mass shift from the conventional one,
    \item Axion DM produiced from PQ Higgs condensate fragmentation where a very strong gravitational wave from the out-of-scaling high-tension string is predicted,
    \item PQ Higgs as a dominant DM through the thermal misalignment
    \item Axion-PQ Higgs co-DM through the stochastic and thermal misalignment.
\end{itemize}

Experimental and observational opportunities were also explored. The detection of a fifth force, axion and PQ Higgs DM signals, and searches for the exotic particles at accelerators are highlighted as key avenues for testing this model. Gravitational wave observations provide another promising direction, offering complementary insights into how the DM is produced in the early Universe.

So far in this paper, I have discussed the simplest case of the FIPQ framework. However, the framework can be extended in various directions. 

\paragraph{DFSZ realization.}
One very straightforward extension is to consider the DFSZ realization. 
In this case, the (a) scenarios should be excluded due to the domain wall problem, but the (Ib) and (IIb) scenarios remain largely unchanged. 
Although in scenario (Ib) we may encounter metastable cosmic strings when the temperature-dependent mass is significant, these strings only form small loops that collapse until the vacuum mass becomes dominant (this can be seen from the left panel of Fig.\ref{fig:lat} where $\Re\F$ does not look the same as $\Im \F$ at the energy equlitbirum, meaning that the vacuum is in the specific direction of $\arg\F$.). The Universe is then filled with a single domain, ensuring the absence of a domain wall problem. 

An interesting prediction is the second Higgs doublet, which may be very heavy in the original PQ model. In FIPQ realization to evade the fine-tuning to the EW scale, the mass scale cannot be much larger than $\L. $ This may be probed in the HL-LHC and future colliders.
\\

If we relax the assumption that all parameters other than the wave-function renormalization constant are of $\mathcal{O}(1)$ in units of $\L$ in the original basis and allow $\L$ to deviate from the EW scale, richer and more interesting phenomena can emerge.

\paragraph{Suppression of the misalignment mechanism for the axion.}
If $\L$ is taken to be sufficiently small, comparable to the QCD scale, such that $m_s$ is as small as the axion mass, the PQ phase transition could occur simultaneously with the QCD phase transition. This would be driven by the thermal effects of massless PQ quarks in the PQ symmetric phase. In such a case, $\F$ would roll toward the strong CP-conserving minimum. DM could then be produced by the coherent oscillation of $s$  and afterward dynamics, while the misalignment mechanism for the QCD axion would be suppressed. \\

Our discussion can also be extended to a generic dark Higgs field associated with either global or gauged symmetries. By making the field weakly coupled, we can alleviate both the quality and hierarchy problems for such symmetries while allowing low mass scales.

\paragraph{Broken phase sphalerons in the fat string core.}
Depending on the specific model, a very fat cosmic string with a core radius $R_s \sim 1/m_s$ may form. In this case, the EW symmetry could be restored within the core, allowing sphaleron effects to become active, while outside the core the EW symmetry remains broken~\cite{Cline:1998rc}. Given the portal coupling, this is possible when $\lambda_P f_a^2 \lesssim - (100\GEV)^2$. Consequently, we might expect an effective sphaleron contribution even at low cosmic temperatures $T < 100\GEV$:
\beq
n_{\rm sph} \sim \frac{R_s^2 \pi \times H^{-1}}{H^{-3}} \sim 5\times 10^{-5} \(\frac{T_R}{50\GEV}\)^4 \(\frac{1\rm m\EV}{m_\F}\)^2.
\eeq
This could contribute to the baryon number violation required for baryogenesis (e.g., producing the baryon-to-entropy ratio $\mathcal{O}(10^{-10})$). Although a no-go theorem was discussed in Ref.~\cite{Cline:1998rc} for a similar scenario, it implicitly assumed that the string core size is around the EW scale inverse. If the string is significantly fatter and phase transition happens late, i.e., with $Z$ much larger than unity, sufficient volume for baryon number generation could be achieved even in the broken phase.

\paragraph{Late-time cosmology with a feebly interacting dark Higgs field.}
For even lighter dark sector fields, small cosmic string loops could induce interesting late-time phenomena. Because these dark sector fields are extremely light, the thermally mass of the dark Higgs field, controlled by the dark sector temperature $T_{\rm dark}$ (which is smaller than the neutrino temperature), dominates the dynamics for the cosmological evolution. 
Since the tensions redshift, they are not harmful. In particular in (Ib) case with oscillation happens around the present Universe, 
fat strings in small loops, potentially attached to domain walls, could be abundant in the Universe. These structures may contribute to cosmic birefringence~\cite{Agrawal:2019lkr, Takahashi:2020tqv, Yin:2021kmx, Kitajima:2022jzz, Gonzalez:2022mcx,Jain:2022jrp, Kitajima:2023kzu} and could also be probed by experiments searching for domain walls~\cite{Masia-Roig:2019hsy,GNOME:2023rpz} on Earth. 
Further study is warranted.

\section*{Acknowledgments}
This work was supported by JSPS KAKENHI Grant No. 20H05851, 21K20364, 22K14029, 22H01215,  and Incentive Research Fund for Young Researchers from Tokyo Metropolitan University. 

\bibliography{reference}

\providecommand{\href}[2]{#2}\begingroup\raggedright\begin{thebibliography}{100}

\bibitem{ParticleDataGroup:2020ssz}
{\scshape Particle Data Group} collaboration, P.~A. Zyla et~al., \emph{{Review
  of Particle Physics}},
  \href{https://doi.org/10.1093/ptep/ptaa104}{\emph{PTEP} {\bfseries 2020}
  (2020) 083C01}.

\bibitem{Peccei:1977hh}
R.~D. Peccei and H.~R. Quinn, \emph{{CP Conservation in the Presence of
  Instantons}}, \href{https://doi.org/10.1103/PhysRevLett.38.1440}{\emph{Phys.
  Rev. Lett.} {\bfseries 38} (1977) 1440}.

\bibitem{Peccei:1977ur}
R.~D. Peccei and H.~R. Quinn, \emph{{Constraints Imposed by CP Conservation in
  the Presence of Instantons}},
  \href{https://doi.org/10.1103/PhysRevD.16.1791}{\emph{Phys. Rev. D}
  {\bfseries 16} (1977) 1791}.

\bibitem{Weinberg:1977ma}
S.~Weinberg, \emph{{A New Light Boson?}},
  \href{https://doi.org/10.1103/PhysRevLett.40.223}{\emph{Phys. Rev. Lett.}
  {\bfseries 40} (1978) 223}.

\bibitem{Wilczek:1977pj}
F.~Wilczek, \emph{{Problem of Strong $P$ and $T$ Invariance in the Presence of
  Instantons}}, \href{https://doi.org/10.1103/PhysRevLett.40.279}{\emph{Phys.
  Rev. Lett.} {\bfseries 40} (1978) 279}.

\bibitem{Mayle:1987as}
R.~Mayle, J.~R. Wilson, J.~R. Ellis, K.~A. Olive, D.~N. Schramm and
  G.~Steigman, \emph{{Constraints on Axions from SN 1987a}},
  \href{https://doi.org/10.1016/0370-2693(88)91595-X}{\emph{Phys. Lett. B}
  {\bfseries 203} (1988) 188}.

\bibitem{Raffelt:1987yt}
G.~Raffelt and D.~Seckel, \emph{{Bounds on Exotic Particle Interactions from SN
  1987a}}, \href{https://doi.org/10.1103/PhysRevLett.60.1793}{\emph{Phys. Rev.
  Lett.} {\bfseries 60} (1988) 1793}.

\bibitem{Chang:2018rso}
J.~H. Chang, R.~Essig and S.~D. McDermott, \emph{{Supernova 1987A Constraints
  on Sub-GeV Dark Sectors, Millicharged Particles, the QCD Axion, and an
  Axion-like Particle}},
  \href{https://doi.org/10.1007/JHEP09(2018)051}{\emph{JHEP} {\bfseries 09}
  (2018) 051} [\href{https://arxiv.org/abs/1803.00993}{{\ttfamily
  1803.00993}}].

\bibitem{Preskill:1982cy}
J.~Preskill, M.~B. Wise and F.~Wilczek, \emph{{Cosmology of the Invisible
  Axion}}, \href{https://doi.org/10.1016/0370-2693(83)90637-8}{\emph{Phys.
  Lett. B} {\bfseries 120} (1983) 127}.

\bibitem{Abbott:1982af}
L.~F. Abbott and P.~Sikivie, \emph{{A Cosmological Bound on the Invisible
  Axion}}, \href{https://doi.org/10.1016/0370-2693(83)90638-X}{\emph{Phys.
  Lett. B} {\bfseries 120} (1983) 133}.

\bibitem{Dine:1982ah}
M.~Dine and W.~Fischler, \emph{{The Not So Harmless Axion}},
  \href{https://doi.org/10.1016/0370-2693(83)90639-1}{\emph{Phys. Lett. B}
  {\bfseries 120} (1983) 137}.

\bibitem{Graham:2018jyp}
P.~W. Graham and A.~Scherlis, \emph{{Stochastic axion scenario}},
  \href{https://doi.org/10.1103/PhysRevD.98.035017}{\emph{Phys. Rev. D}
  {\bfseries 98} (2018) 035017}
  [\href{https://arxiv.org/abs/1805.07362}{{\ttfamily 1805.07362}}].

\bibitem{Guth:2018hsa}
F.~Takahashi, W.~Yin and A.~H. Guth, \emph{{QCD axion window and low-scale
  inflation}}, \href{https://doi.org/10.1103/PhysRevD.98.015042}{\emph{Phys.
  Rev. D} {\bfseries 98} (2018) 015042}
  [\href{https://arxiv.org/abs/1805.08763}{{\ttfamily 1805.08763}}].

\bibitem{Kim:1979if}
J.~E. Kim, \emph{{Weak Interaction Singlet and Strong CP Invariance}},
  \href{https://doi.org/10.1103/PhysRevLett.43.103}{\emph{Phys. Rev. Lett.}
  {\bfseries 43} (1979) 103}.

\bibitem{Shifman:1979if}
M.~A. Shifman, A.~I. Vainshtein and V.~I. Zakharov, \emph{{Can Confinement
  Ensure Natural CP Invariance of Strong Interactions?}},
  \href{https://doi.org/10.1016/0550-3213(80)90209-6}{\emph{Nucl. Phys. B}
  {\bfseries 166} (1980) 493}.

\bibitem{Dine:1981rt}
M.~Dine, W.~Fischler and M.~Srednicki, \emph{{A Simple Solution to the Strong
  CP Problem with a Harmless Axion}},
  \href{https://doi.org/10.1016/0370-2693(81)90590-6}{\emph{Phys. Lett. B}
  {\bfseries 104} (1981) 199}.

\bibitem{Zhitnitsky:1980tq}
A.~R. Zhitnitsky, \emph{{On Possible Suppression of the Axion Hadron
  Interactions. (In Russian)}}, {\emph{Sov. J. Nucl. Phys.} {\bfseries 31}
  (1980) 260}.

\bibitem{Saikawa:2024bta}
K.~Saikawa, J.~Redondo, A.~Vaquero and M.~Kaltschmidt, \emph{{Spectrum of
  global string networks and the axion dark matter mass}},
  \href{https://doi.org/10.1088/1475-7516/2024/10/043}{\emph{JCAP} {\bfseries
  10} (2024) 043} [\href{https://arxiv.org/abs/2401.17253}{{\ttfamily
  2401.17253}}].

\bibitem{Kim:2024wku}
H.~Kim, J.~Park and M.~Son, \emph{{Axion dark matter from cosmic string
  network}}, \href{https://doi.org/10.1007/JHEP07(2024)150}{\emph{JHEP}
  {\bfseries 07} (2024) 150}
  [\href{https://arxiv.org/abs/2402.00741}{{\ttfamily 2402.00741}}].

\bibitem{Buschmann:2024bfj}
M.~Buschmann, \emph{{Sledgehamr: Simulating Scalar Fields with Adaptive Mesh
  Refinement}},  \href{https://arxiv.org/abs/2404.02950}{{\ttfamily
  2404.02950}}.

\bibitem{Benabou:2024msj}
J.~N. Benabou, M.~Buschmann, J.~W. Foster and B.~R. Safdi, \emph{{Axion mass
  prediction from adaptive mesh refinement cosmological lattice simulations}},
  \href{https://arxiv.org/abs/2412.08699}{{\ttfamily 2412.08699}}.

\bibitem{Kim:2024dtq}
H.~Kim and M.~Son, \emph{{More Scalings from Cosmic Strings}},
  \href{https://arxiv.org/abs/2411.08455}{{\ttfamily 2411.08455}}.

\bibitem{Sikivie:1982qv}
P.~Sikivie, \emph{{Of Axions, Domain Walls and the Early Universe}},
  \href{https://doi.org/10.1103/PhysRevLett.48.1156}{\emph{Phys. Rev. Lett.}
  {\bfseries 48} (1982) 1156}.

\bibitem{Vilenkin:1982ks}
A.~Vilenkin and A.~E. Everett, \emph{{Cosmic Strings and Domain Walls in Models
  with Goldstone and PseudoGoldstone Bosons}},
  \href{https://doi.org/10.1103/PhysRevLett.48.1867}{\emph{Phys. Rev. Lett.}
  {\bfseries 48} (1982) 1867}.

\bibitem{Harari:1987ht}
D.~Harari and P.~Sikivie, \emph{{On the Evolution of Global Strings in the
  Early Universe}},
  \href{https://doi.org/10.1016/0370-2693(87)90032-3}{\emph{Phys. Lett. B}
  {\bfseries 195} (1987) 361}.

\bibitem{Davis:1986xc}
R.~L. Davis, \emph{{Cosmic Axions from Cosmic Strings}},
  \href{https://doi.org/10.1016/0370-2693(86)90300-X}{\emph{Phys. Lett. B}
  {\bfseries 180} (1986) 225}.

\bibitem{Dine:2020pds}
M.~Dine, N.~Fernandez, A.~Ghalsasi and H.~H. Patel, \emph{{Comments on axions,
  domain walls, and cosmic strings}},
  \href{https://doi.org/10.1088/1475-7516/2021/11/041}{\emph{JCAP} {\bfseries
  11} (2021) 041} [\href{https://arxiv.org/abs/2012.13065}{{\ttfamily
  2012.13065}}].

\bibitem{Batell:2021ofv}
B.~Batell and A.~Ghalsasi, \emph{{Thermal misalignment of scalar dark matter}},
  \href{https://doi.org/10.1103/PhysRevD.107.L091701}{\emph{Phys. Rev. D}
  {\bfseries 107} (2023) L091701}
  [\href{https://arxiv.org/abs/2109.04476}{{\ttfamily 2109.04476}}].

\bibitem{Batell:2022qvr}
B.~Batell, A.~Ghalsasi and M.~Rai, \emph{{Dynamics of dark matter misalignment
  through the Higgs portal}},
  \href{https://doi.org/10.1007/JHEP01(2024)038}{\emph{JHEP} {\bfseries 01}
  (2024) 038} [\href{https://arxiv.org/abs/2211.09132}{{\ttfamily
  2211.09132}}].

\bibitem{Choi:2015fiu}
K.~Choi and S.~H. Im, \emph{{Realizing the relaxion from multiple axions and
  its UV completion with high scale supersymmetry}},
  \href{https://doi.org/10.1007/JHEP01(2016)149}{\emph{JHEP} {\bfseries 01}
  (2016) 149} [\href{https://arxiv.org/abs/1511.00132}{{\ttfamily
  1511.00132}}].

\bibitem{Kaplan:2015fuy}
D.~E. Kaplan and R.~Rattazzi, \emph{{Large field excursions and approximate
  discrete symmetries from a clockwork axion}},
  \href{https://doi.org/10.1103/PhysRevD.93.085007}{\emph{Phys. Rev. D}
  {\bfseries 93} (2016) 085007}
  [\href{https://arxiv.org/abs/1511.01827}{{\ttfamily 1511.01827}}].

\bibitem{Giudice:2016yja}
G.~F. Giudice and M.~McCullough, \emph{{A Clockwork Theory}},
  \href{https://doi.org/10.1007/JHEP02(2017)036}{\emph{JHEP} {\bfseries 02}
  (2017) 036} [\href{https://arxiv.org/abs/1610.07962}{{\ttfamily
  1610.07962}}].

\bibitem{Higaki:2016yqk}
T.~Higaki, K.~S. Jeong, N.~Kitajima and F.~Takahashi, \emph{{Quality of the
  Peccei-Quinn symmetry in the Aligned QCD Axion and Cosmological
  Implications}}, \href{https://doi.org/10.1007/JHEP06(2016)150}{\emph{JHEP}
  {\bfseries 06} (2016) 150}
  [\href{https://arxiv.org/abs/1603.02090}{{\ttfamily 1603.02090}}].

\bibitem{CMS:2019afi}
{\scshape CMS} collaboration, A.~M. Sirunyan et~al., \emph{{Search for
  electroweak production of a vector-like T quark using fully hadronic final
  states}}, \href{https://doi.org/10.1007/JHEP01(2020)036}{\emph{JHEP}
  {\bfseries 01} (2020) 036}
  [\href{https://arxiv.org/abs/1909.04721}{{\ttfamily 1909.04721}}].

\bibitem{ATLAS:2022ozf}
{\scshape ATLAS} collaboration, G.~Aad et~al., \emph{{Search for single
  production of a vectorlike $T$ quark decaying into a Higgs boson and top
  quark with fully hadronic final states using the ATLAS detector}},
  \href{https://doi.org/10.1103/PhysRevD.105.092012}{\emph{Phys. Rev. D}
  {\bfseries 105} (2022) 092012}
  [\href{https://arxiv.org/abs/2201.07045}{{\ttfamily 2201.07045}}].

\bibitem{tHooft:1979rat}
G.~'t~Hooft, \emph{{Naturalness, chiral symmetry, and spontaneous chiral
  symmetry breaking}},
  \href{https://doi.org/10.1007/978-1-4684-7571-5_9}{\emph{NATO Sci. Ser. B}
  {\bfseries 59} (1980) 135}.

\bibitem{Piazza:2010ye}
F.~Piazza and M.~Pospelov, \emph{{Sub-eV scalar dark matter through the
  super-renormalizable Higgs portal}},
  \href{https://doi.org/10.1103/PhysRevD.82.043533}{\emph{Phys. Rev. D}
  {\bfseries 82} (2010) 043533}
  [\href{https://arxiv.org/abs/1003.2313}{{\ttfamily 1003.2313}}].

\bibitem{Leutwyler:1989tn}
H.~Leutwyler and M.~A. Shifman, \emph{{GOLDSTONE BOSONS GENERATE PECULIAR
  CONFORMAL ANOMALIES}},
  \href{https://doi.org/10.1016/0370-2693(89)91730-9}{\emph{Phys. Lett. B}
  {\bfseries 221} (1989) 384}.

\bibitem{NA62:2021zjw}
{\scshape NA62} collaboration, E.~Cortina~Gil et~al., \emph{{Measurement of the
  very rare K$^{+}$\textrightarrow{}$ {\pi}^{+}\nu \overline{\nu} $ decay}},
  \href{https://doi.org/10.1007/JHEP06(2021)093}{\emph{JHEP} {\bfseries 06}
  (2021) 093} [\href{https://arxiv.org/abs/2103.15389}{{\ttfamily
  2103.15389}}].

\bibitem{NA62:2020pwi}
{\scshape NA62} collaboration, E.~Cortina~Gil et~al., \emph{{Search for $\pi^0$
  decays to invisible particles}},
  \href{https://doi.org/10.1007/JHEP02(2021)201}{\emph{JHEP} {\bfseries 02}
  (2021) 201} [\href{https://arxiv.org/abs/2010.07644}{{\ttfamily
  2010.07644}}].

\bibitem{NA62:2020xlg}
{\scshape NA62} collaboration, E.~Cortina~Gil et~al., \emph{{Search for a
  feebly interacting particle $X$ in the decay $K^{+}\rightarrow\pi^{+}X$}},
  \href{https://doi.org/10.1007/JHEP03(2021)058}{\emph{JHEP} {\bfseries 03}
  (2021) 058} [\href{https://arxiv.org/abs/2011.11329}{{\ttfamily
  2011.11329}}].

\bibitem{CHARM:1985anb}
{\scshape CHARM} collaboration, F.~Bergsma et~al., \emph{{Search for Axion Like
  Particle Production in 400-{GeV} Proton - Copper Interactions}},
  \href{https://doi.org/10.1016/0370-2693(85)90400-9}{\emph{Phys. Lett. B}
  {\bfseries 157} (1985) 458}.

\bibitem{LHCb:2012juf}
{\scshape LHCb} collaboration, R.~Aaij et~al., \emph{{Differential branching
  fraction and angular analysis of the $B^{+} \rightarrow K^{+}\mu^{+}\mu^{-}$
  decay}}, \href{https://doi.org/10.1007/JHEP02(2013)105}{\emph{JHEP}
  {\bfseries 02} (2013) 105} [\href{https://arxiv.org/abs/1209.4284}{{\ttfamily
  1209.4284}}].

\bibitem{L3:1996ome}
{\scshape L3} collaboration, M.~Acciarri et~al., \emph{{Search for neutral
  Higgs boson production through the process e+ e- --\ensuremath{>} Z* H0}},
  \href{https://doi.org/10.1016/0370-2693(96)00987-2}{\emph{Phys. Lett. B}
  {\bfseries 385} (1996) 454}.

\bibitem{Dev:2020eam}
P.~S.~B. Dev, R.~N. Mohapatra and Y.~Zhang, \emph{{Revisiting supernova
  constraints on a light CP-even scalar}},
  \href{https://doi.org/10.1088/1475-7516/2020/08/003}{\emph{JCAP} {\bfseries
  08} (2020) 003} [\href{https://arxiv.org/abs/2005.00490}{{\ttfamily
  2005.00490}}].

\bibitem{Salumbides:2013dua}
E.~J. Salumbides, W.~Ubachs and V.~I. Korobov, \emph{{Bounds on fifth forces at
  the sub-Angstrom length scale}},
  \href{https://doi.org/10.1016/j.jms.2014.04.003}{\emph{J. Molec. Spectrosc.}
  {\bfseries 300} (2014) 65} [\href{https://arxiv.org/abs/1308.1711}{{\ttfamily
  1308.1711}}].

\bibitem{Adelberger:2009zz}
E.~G. Adelberger, J.~H. Gundlach, B.~R. Heckel, S.~Hoedl and S.~Schlamminger,
  \emph{{Torsion balance experiments: A low-energy frontier of particle
  physics}}, \href{https://doi.org/10.1016/j.ppnp.2008.08.002}{\emph{Prog.
  Part. Nucl. Phys.} {\bfseries 62} (2009) 102}.

\bibitem{Salumbides:2013aga}
E.~J. Salumbides, J.~C.~J. Koelemeij, J.~Komasa, K.~Pachucki, K.~S.~E. Eikema
  and W.~Ubachs, \emph{{Bounds on fifth forces from precision measurements on
  molecules}}, \href{https://doi.org/10.1103/PhysRevD.87.112008}{\emph{Phys.
  Rev. D} {\bfseries 87} (2013) 112008}
  [\href{https://arxiv.org/abs/1304.6560}{{\ttfamily 1304.6560}}].

\bibitem{Yin:2024lla}
W.~Yin et~al., \emph{{First Result for Dark Matter Search by WINERED}},
  \href{https://arxiv.org/abs/2402.07976}{{\ttfamily 2402.07976}}.

\bibitem{Wadekar:2021qae}
D.~Wadekar and Z.~Wang, \emph{{Strong constraints on decay and annihilation of
  dark matter from heating of gas-rich dwarf galaxies}},
  \href{https://doi.org/10.1103/PhysRevD.106.075007}{\emph{Phys. Rev. D}
  {\bfseries 106} (2022) 075007}
  [\href{https://arxiv.org/abs/2111.08025}{{\ttfamily 2111.08025}}].

\bibitem{Janish:2023kvi}
R.~Janish and E.~Pinetti, \emph{{Hunting Dark Matter Lines in the Infrared
  Background with the James Webb Space Telescope}},
  \href{https://arxiv.org/abs/2310.15395}{{\ttfamily 2310.15395}}.

\bibitem{Calore:2022pks}
F.~Calore, A.~Dekker, P.~D. Serpico and T.~Siegert, \emph{{Constraints on light
  decaying dark matter candidates from 16~yr of INTEGRAL/SPI observations}},
  \href{https://doi.org/10.1093/mnras/stad457}{\emph{Mon. Not. Roy. Astron.
  Soc.} {\bfseries 520} (2023) 4167}
  [\href{https://arxiv.org/abs/2209.06299}{{\ttfamily 2209.06299}}].

\bibitem{Sun:2023acy}
Y.~Sun, J.~W. Foster, H.~Liu, J.~B. Mu\~noz and T.~R. Slatyer,
  \emph{{Inhomogeneous Energy Injection in the 21-cm Power Spectrum:
  Sensitivity to Dark Matter Decay}},
  \href{https://arxiv.org/abs/2312.11608}{{\ttfamily 2312.11608}}.

\bibitem{Caputo:2022mah}
A.~Caputo, H.-T. Janka, G.~Raffelt and E.~Vitagliano, \emph{{Low-Energy
  Supernovae Severely Constrain Radiative Particle Decays}},
  \href{https://doi.org/10.1103/PhysRevLett.128.221103}{\emph{Phys. Rev. Lett.}
  {\bfseries 128} (2022) 221103}
  [\href{https://arxiv.org/abs/2201.09890}{{\ttfamily 2201.09890}}].

\bibitem{Nakayama:2022jza}
K.~Nakayama and W.~Yin, \emph{{Anisotropic cosmic optical background bound for
  decaying dark matter in light of the LORRI anomaly}},
  \href{https://doi.org/10.1103/PhysRevD.106.103505}{\emph{Phys. Rev. D}
  {\bfseries 106} (2022) 103505}
  [\href{https://arxiv.org/abs/2205.01079}{{\ttfamily 2205.01079}}].

\bibitem{Carenza:2023qxh}
P.~Carenza, G.~Lucente and E.~Vitagliano, \emph{{Probing the blue axion with
  cosmic optical background anisotropies}},
  \href{https://doi.org/10.1103/PhysRevD.107.083032}{\emph{Phys. Rev. D}
  {\bfseries 107} (2023) 083032}
  [\href{https://arxiv.org/abs/2301.06560}{{\ttfamily 2301.06560}}].

\bibitem{Sakurai:2021ipp}
K.~Sakurai and W.~Yin, \emph{{Phenomenology of CP-even ALP}},
  \href{https://doi.org/10.1007/JHEP04(2022)113}{\emph{JHEP} {\bfseries 04}
  (2022) 113} [\href{https://arxiv.org/abs/2111.03653}{{\ttfamily
  2111.03653}}].

\bibitem{AxionLimits}
C.~O'Hare, ``cajohare/axionlimits: Axionlimits.''
  \url{https://cajohare.github.io/AxionLimits/}, 2020.
\newblock 10.5281/zenodo.3932430.

\bibitem{Haghighat:2022qyh}
G.~Haghighat, M.~Mohammadi~Najafabadi, K.~Sakurai and W.~Yin, \emph{{Probing a
  light dark sector at future lepton colliders via invisible decays of the
  SM-like and dark Higgs bosons}},
  \href{https://doi.org/10.1103/PhysRevD.107.035033}{\emph{Phys. Rev. D}
  {\bfseries 107} (2023) 035033}
  [\href{https://arxiv.org/abs/2209.07565}{{\ttfamily 2209.07565}}].

\bibitem{Lee:1988ge}
K.-M. Lee, \emph{{Wormholes and Goldstone Bosons}},
  \href{https://doi.org/10.1103/PhysRevLett.61.263}{\emph{Phys. Rev. Lett.}
  {\bfseries 61} (1988) 263}.

\bibitem{Giddings:1987cg}
S.~B. Giddings and A.~Strominger, \emph{{Axion Induced Topology Change in
  Quantum Gravity and String Theory}},
  \href{https://doi.org/10.1016/0550-3213(88)90446-4}{\emph{Nucl. Phys. B}
  {\bfseries 306} (1988) 890}.

\bibitem{Abbott:1989jw}
L.~F. Abbott and M.~B. Wise, \emph{{Wormholes and Global Symmetries}},
  \href{https://doi.org/10.1016/0550-3213(89)90503-8}{\emph{Nucl. Phys. B}
  {\bfseries 325} (1989) 687}.

\bibitem{Kallosh:1995hi}
R.~Kallosh, A.~D. Linde, D.~A. Linde and L.~Susskind, \emph{{Gravity and global
  symmetries}}, \href{https://doi.org/10.1103/PhysRevD.52.912}{\emph{Phys. Rev.
  D} {\bfseries 52} (1995) 912}
  [\href{https://arxiv.org/abs/hep-th/9502069}{{\ttfamily hep-th/9502069}}].

\bibitem{Alvey:2020nyh}
J.~Alvey and M.~Escudero, \emph{{The axion quality problem: global symmetry
  breaking and wormholes}},
  \href{https://doi.org/10.1007/JHEP01(2021)032}{\emph{JHEP} {\bfseries 01}
  (2021) 032} [\href{https://arxiv.org/abs/2009.03917}{{\ttfamily
  2009.03917}}].

\bibitem{Hamaguchi:2021mmt}
K.~Hamaguchi, Y.~Kanazawa and N.~Nagata, \emph{{Axion quality problem
  alleviated by nonminimal coupling to gravity}},
  \href{https://doi.org/10.1103/PhysRevD.105.076008}{\emph{Phys. Rev. D}
  {\bfseries 105} (2022) 076008}
  [\href{https://arxiv.org/abs/2108.13245}{{\ttfamily 2108.13245}}].

\bibitem{Dine:1986bg}
M.~Dine and N.~Seiberg, \emph{{String Theory and the Strong {CP} Problem}},
  \href{https://doi.org/10.1016/0550-3213(86)90043-X}{\emph{Nucl. Phys. B}
  {\bfseries 273} (1986) 109}.

\bibitem{Kitano:2021fdl}
R.~Kitano and W.~Yin, \emph{{Strong CP problem and axion dark matter with small
  instantons}}, \href{https://doi.org/10.1007/JHEP07(2021)078}{\emph{JHEP}
  {\bfseries 07} (2021) 078}
  [\href{https://arxiv.org/abs/2103.08598}{{\ttfamily 2103.08598}}].

\bibitem{Poppitz:2002ac}
E.~Poppitz and Y.~Shirman, \emph{{The Strength of small instanton amplitudes in
  gauge theories with compact extra dimensions}},
  \href{https://doi.org/10.1088/1126-6708/2002/07/041}{\emph{JHEP} {\bfseries
  07} (2002) 041} [\href{https://arxiv.org/abs/hep-th/0204075}{{\ttfamily
  hep-th/0204075}}].

\bibitem{Gherghetta:2020keg}
T.~Gherghetta, V.~V. Khoze, A.~Pomarol and Y.~Shirman, \emph{{The Axion Mass
  from 5D Small Instantons}},
  \href{https://doi.org/10.1007/JHEP03(2020)063}{\emph{JHEP} {\bfseries 03}
  (2020) 063} [\href{https://arxiv.org/abs/2001.05610}{{\ttfamily
  2001.05610}}].

\bibitem{Agrawal:2017ksf}
P.~Agrawal and K.~Howe, \emph{{Factoring the Strong CP Problem}},
  \href{https://doi.org/10.1007/JHEP12(2018)029}{\emph{JHEP} {\bfseries 12}
  (2018) 029} [\href{https://arxiv.org/abs/1710.04213}{{\ttfamily
  1710.04213}}].

\bibitem{Agrawal:2017evu}
P.~Agrawal and K.~Howe, \emph{{A Flavorful Factoring of the Strong CP
  Problem}}, \href{https://doi.org/10.1007/JHEP12(2018)035}{\emph{JHEP}
  {\bfseries 12} (2018) 035}
  [\href{https://arxiv.org/abs/1712.05803}{{\ttfamily 1712.05803}}].

\bibitem{Fuentes-Martin:2019bue}
J.~Fuentes-Mart\'\i{}n, M.~Reig and A.~Vicente, \emph{{Strong $CP$ problem with
  low-energy emergent QCD: The 4321 case}},
  \href{https://doi.org/10.1103/PhysRevD.100.115028}{\emph{Phys. Rev. D}
  {\bfseries 100} (2019) 115028}
  [\href{https://arxiv.org/abs/1907.02550}{{\ttfamily 1907.02550}}].

\bibitem{Csaki:2019vte}
C.~Cs\'aki, M.~Ruhdorfer and Y.~Shirman, \emph{{UV Sensitivity of the Axion
  Mass from Instantons in Partially Broken Gauge Groups}},
  \href{https://doi.org/10.1007/JHEP04(2020)031}{\emph{JHEP} {\bfseries 04}
  (2020) 031} [\href{https://arxiv.org/abs/1912.02197}{{\ttfamily
  1912.02197}}].

\bibitem{Takahashi:2021tff}
F.~Takahashi and W.~Yin, \emph{{Challenges for heavy QCD axion inflation}},
  \href{https://doi.org/10.1088/1475-7516/2021/10/057}{\emph{JCAP} {\bfseries
  10} (2021) 057} [\href{https://arxiv.org/abs/2105.10493}{{\ttfamily
  2105.10493}}].

\bibitem{Babu:2024udi}
K.~S. Babu, B.~Dutta and R.~N. Mohapatra, \emph{{Hybrid SO(10) Axion Model
  Without Quality Problem}},
  \href{https://arxiv.org/abs/2410.07323}{{\ttfamily 2410.07323}}.

\bibitem{Gherghetta:2020ofz}
T.~Gherghetta and M.~D. Nguyen, \emph{{A Composite Higgs with a Heavy Composite
  Axion}}, \href{https://doi.org/10.1007/JHEP12(2020)094}{\emph{JHEP}
  {\bfseries 12} (2020) 094}
  [\href{https://arxiv.org/abs/2007.10875}{{\ttfamily 2007.10875}}].

\bibitem{Aoki:2024usv}
T.~Aoki, M.~Ibe, S.~Shirai and K.~Watanabe, \emph{{Small instanton effects on
  composite axion mass}},
  \href{https://doi.org/10.1007/JHEP07(2024)269}{\emph{JHEP} {\bfseries 07}
  (2024) 269} [\href{https://arxiv.org/abs/2404.19342}{{\ttfamily
  2404.19342}}].

\bibitem{Wada:2024txn}
J.~Wada and W.~Yin, \emph{{Gauge coupling jump and small instantons from a
  large non-minimal coupling}},
  \href{https://arxiv.org/abs/2411.00768}{{\ttfamily 2411.00768}}.

\bibitem{Bezrukov:2007ep}
F.~L. Bezrukov and M.~Shaposhnikov, \emph{{The Standard Model Higgs boson as
  the inflaton}},
  \href{https://doi.org/10.1016/j.physletb.2007.11.072}{\emph{Phys. Lett. B}
  {\bfseries 659} (2008) 703}
  [\href{https://arxiv.org/abs/0710.3755}{{\ttfamily 0710.3755}}].

\bibitem{Bezrukov:2008ej}
F.~L. Bezrukov, A.~Magnin and M.~Shaposhnikov, \emph{{Standard Model Higgs
  boson mass from inflation}},
  \href{https://doi.org/10.1016/j.physletb.2009.03.035}{\emph{Phys. Lett. B}
  {\bfseries 675} (2009) 88} [\href{https://arxiv.org/abs/0812.4950}{{\ttfamily
  0812.4950}}].

\bibitem{Takahashi:2019qmh}
F.~Takahashi and W.~Yin, \emph{{ALP inflation and Big Bang on Earth}},
  \href{https://doi.org/10.1007/JHEP07(2019)095}{\emph{JHEP} {\bfseries 07}
  (2019) 095} [\href{https://arxiv.org/abs/1903.00462}{{\ttfamily
  1903.00462}}].

\bibitem{Yin:2022fgo}
W.~Yin, \emph{{Weak-scale Higgs inflation}},
  \href{https://doi.org/10.1088/1475-7516/2024/05/060}{\emph{JCAP} {\bfseries
  05} (2024) 060} [\href{https://arxiv.org/abs/2210.15680}{{\ttfamily
  2210.15680}}].

\bibitem{Jaeckel:2020oet}
J.~Jaeckel and W.~Yin, \emph{{Boosted Neutrinos and Relativistic Dark Particles
  as Messengers from Reheating}},
  \href{https://doi.org/10.1088/1475-7516/2021/02/044}{\emph{JCAP} {\bfseries
  02} (2021) 044} [\href{https://arxiv.org/abs/2007.15006}{{\ttfamily
  2007.15006}}].

\bibitem{Starobinsky:1986fx}
A.~A. Starobinsky, \emph{{STOCHASTIC DE SITTER (INFLATIONARY) STAGE IN THE
  EARLY UNIVERSE}}, \href{https://doi.org/10.1007/3-540-16452-9_6}{\emph{Lect.
  Notes Phys.} {\bfseries 246} (1986) 107}.

\bibitem{Starobinsky:1994bd}
A.~A. Starobinsky and J.~Yokoyama, \emph{{Equilibrium state of a
  selfinteracting scalar field in the De Sitter background}},
  \href{https://doi.org/10.1103/PhysRevD.50.6357}{\emph{Phys. Rev. D}
  {\bfseries 50} (1994) 6357}
  [\href{https://arxiv.org/abs/astro-ph/9407016}{{\ttfamily
  astro-ph/9407016}}].

\bibitem{Hardwick:2017fjo}
R.~J. Hardwick, V.~Vennin, C.~T. Byrnes, J.~Torrado and D.~Wands, \emph{{The
  stochastic spectator}},
  \href{https://doi.org/10.1088/1475-7516/2017/10/018}{\emph{JCAP} {\bfseries
  10} (2017) 018} [\href{https://arxiv.org/abs/1701.06473}{{\ttfamily
  1701.06473}}].

\bibitem{Ho:2019ayl}
S.-Y. Ho, F.~Takahashi and W.~Yin, \emph{{Relaxing the Cosmological Moduli
  Problem by Low-scale Inflation}},
  \href{https://doi.org/10.1007/JHEP04(2019)149}{\emph{JHEP} {\bfseries 04}
  (2019) 149} [\href{https://arxiv.org/abs/1901.01240}{{\ttfamily
  1901.01240}}].

\bibitem{Alonso-Alvarez:2019ixv}
G.~Alonso-\'Alvarez, T.~Hugle and J.~Jaeckel, \emph{{Misalignment
  \textbackslash{}\& Co.: (Pseudo-)scalar and vector dark matter with curvature
  couplings}}, \href{https://doi.org/10.1088/1475-7516/2020/02/014}{\emph{JCAP}
  {\bfseries 02} (2020) 014}
  [\href{https://arxiv.org/abs/1905.09836}{{\ttfamily 1905.09836}}].

\bibitem{Planck:2018jri}
{\scshape Planck} collaboration, Y.~Akrami et~al., \emph{{Planck 2018 results.
  X. Constraints on inflation}},
  \href{https://doi.org/10.1051/0004-6361/201833887}{\emph{Astron. Astrophys.}
  {\bfseries 641} (2020) A10}
  [\href{https://arxiv.org/abs/1807.06211}{{\ttfamily 1807.06211}}].

\bibitem{BICEP:2021xfz}
{\scshape BICEP, Keck} collaboration, P.~A.~R. Ade et~al., \emph{{Improved
  Constraints on Primordial Gravitational Waves using Planck, WMAP, and
  BICEP/Keck Observations through the 2018 Observing Season}},
  \href{https://doi.org/10.1103/PhysRevLett.127.151301}{\emph{Phys. Rev. Lett.}
  {\bfseries 127} (2021) 151301}
  [\href{https://arxiv.org/abs/2110.00483}{{\ttfamily 2110.00483}}].

\bibitem{Kobayashi:2013nva}
T.~Kobayashi, R.~Kurematsu and F.~Takahashi, \emph{{Isocurvature Constraints
  and Anharmonic Effects on QCD Axion Dark Matter}},
  \href{https://doi.org/10.1088/1475-7516/2013/09/032}{\emph{JCAP} {\bfseries
  09} (2013) 032} [\href{https://arxiv.org/abs/1304.0922}{{\ttfamily
  1304.0922}}].

\bibitem{Nakayama:2021avl}
K.~Nakayama and W.~Yin, \emph{{Hidden photon and axion dark matter from
  symmetry breaking}},
  \href{https://doi.org/10.1007/JHEP10(2021)026}{\emph{JHEP} {\bfseries 10}
  (2021) 026} [\href{https://arxiv.org/abs/2105.14549}{{\ttfamily
  2105.14549}}].

\bibitem{Kawasaki:2017bqm}
M.~Kawasaki, K.~Kohri, T.~Moroi and Y.~Takaesu, \emph{{Revisiting Big-Bang
  Nucleosynthesis Constraints on Long-Lived Decaying Particles}},
  \href{https://doi.org/10.1103/PhysRevD.97.023502}{\emph{Phys. Rev. D}
  {\bfseries 97} (2018) 023502}
  [\href{https://arxiv.org/abs/1709.01211}{{\ttfamily 1709.01211}}].

\bibitem{Poulin:2016anj}
V.~Poulin, J.~Lesgourgues and P.~D. Serpico, \emph{{Cosmological constraints on
  exotic injection of electromagnetic energy}},
  \href{https://doi.org/10.1088/1475-7516/2017/03/043}{\emph{JCAP} {\bfseries
  03} (2017) 043} [\href{https://arxiv.org/abs/1610.10051}{{\ttfamily
  1610.10051}}].

\bibitem{Moroi:2020has}
T.~Moroi and W.~Yin, \emph{{Light Dark Matter from Inflaton Decay}},
  \href{https://doi.org/10.1007/JHEP03(2021)301}{\emph{JHEP} {\bfseries 03}
  (2021) 301} [\href{https://arxiv.org/abs/2011.09475}{{\ttfamily
  2011.09475}}].

\bibitem{Moroi:2020bkq}
T.~Moroi and W.~Yin, \emph{{Particle Production from Oscillating Scalar Field
  and Consistency of Boltzmann Equation}},
  \href{https://doi.org/10.1007/JHEP03(2021)296}{\emph{JHEP} {\bfseries 03}
  (2021) 296} [\href{https://arxiv.org/abs/2011.12285}{{\ttfamily
  2011.12285}}].

\bibitem{Choi:2023jxw}
K.-Y. Choi, J.-O. Gong, J.~Joh, W.-I. Park and O.~Seto, \emph{{Light cold dark
  matter from non-thermal decay}},
  \href{https://doi.org/10.1016/j.physletb.2023.138126}{\emph{Phys. Lett. B}
  {\bfseries 845} (2023) 138126}
  [\href{https://arxiv.org/abs/2304.07462}{{\ttfamily 2304.07462}}].

\bibitem{Yin:2023jjj}
W.~Yin, \emph{{Thermal production of cold \textquotedblleft{}hot dark
  matter\textquotedblright{} around eV}},
  \href{https://doi.org/10.1007/JHEP05(2023)180}{\emph{JHEP} {\bfseries 05}
  (2023) 180} [\href{https://arxiv.org/abs/2301.08735}{{\ttfamily
  2301.08735}}].

\bibitem{Sakurai:2024apm}
K.~Sakurai and W.~Yin, \emph{{Stimulated Emission of Dark Matter via Thermal
  Scattering: Novel Limits for Freeze-In and eV Cold Dark Matter}},
  \href{https://arxiv.org/abs/2410.18968}{{\ttfamily 2410.18968}}.

\bibitem{QUAX:2024fut}
{\scshape QUAX} collaboration, A.~Rettaroli et~al., \emph{{Search for axion
  dark matter with the QUAX\textendash{}LNF tunable haloscope}},
  \href{https://doi.org/10.1103/PhysRevD.110.022008}{\emph{Phys. Rev. D}
  {\bfseries 110} (2024) 022008}
  [\href{https://arxiv.org/abs/2402.19063}{{\ttfamily 2402.19063}}].

\bibitem{Beurthey:2020yuq}
S.~Beurthey et~al., \emph{{MADMAX Status Report}},
  \href{https://arxiv.org/abs/2003.10894}{{\ttfamily 2003.10894}}.

\bibitem{Lawson:2019brd}
M.~Lawson, A.~J. Millar, M.~Pancaldi, E.~Vitagliano and F.~Wilczek,
  \emph{{Tunable axion plasma haloscopes}},
  \href{https://doi.org/10.1103/PhysRevLett.123.141802}{\emph{Phys. Rev. Lett.}
  {\bfseries 123} (2019) 141802}
  [\href{https://arxiv.org/abs/1904.11872}{{\ttfamily 1904.11872}}].

\bibitem{Stern:2016bbw}
I.~Stern, \emph{{ADMX Status}},
  \href{https://doi.org/10.22323/1.282.0198}{\emph{PoS} {\bfseries ICHEP2016}
  (2016) 198} [\href{https://arxiv.org/abs/1612.08296}{{\ttfamily
  1612.08296}}].

\bibitem{Yin:2024trc}
W.~Yin, S.~Nakagawa, T.~Murokoshi and M.~Hattori, \emph{{Asymmetric Warm Dark
  Matter: from Cosmological Asymmetry to Chirality of Life}},
  \href{https://arxiv.org/abs/2405.10303}{{\ttfamily 2405.10303}}.

\bibitem{Felder:2000hj}
G.~N. Felder, J.~Garcia-Bellido, P.~B. Greene, L.~Kofman, A.~D. Linde and
  I.~Tkachev, \emph{{Dynamics of symmetry breaking and tachyonic preheating}},
  \href{https://doi.org/10.1103/PhysRevLett.87.011601}{\emph{Phys. Rev. Lett.}
  {\bfseries 87} (2001) 011601}
  [\href{https://arxiv.org/abs/hep-ph/0012142}{{\ttfamily hep-ph/0012142}}].

\bibitem{Felder:2001kt}
G.~N. Felder, L.~Kofman and A.~D. Linde, \emph{{Tachyonic instability and
  dynamics of spontaneous symmetry breaking}},
  \href{https://doi.org/10.1103/PhysRevD.64.123517}{\emph{Phys. Rev. D}
  {\bfseries 64} (2001) 123517}
  [\href{https://arxiv.org/abs/hep-th/0106179}{{\ttfamily hep-th/0106179}}].

\bibitem{WY}
W.~Yin, \emph{{To appear soon}}, .

\bibitem{Figueroa:2020rrl}
D.~G. Figueroa, A.~Florio, F.~Torrenti and W.~Valkenburg, \emph{{The art of
  simulating the early Universe -- Part I}},
  \href{https://doi.org/10.1088/1475-7516/2021/04/035}{\emph{JCAP} {\bfseries
  04} (2021) 035} [\href{https://arxiv.org/abs/2006.15122}{{\ttfamily
  2006.15122}}].

\bibitem{Figueroa:2021yhd}
D.~G. Figueroa, A.~Florio, F.~Torrenti and W.~Valkenburg, \emph{{CosmoLattice:
  A modern code for lattice simulations of scalar and gauge field dynamics in
  an expanding universe}},
  \href{https://doi.org/10.1016/j.cpc.2022.108586}{\emph{Comput. Phys. Commun.}
  {\bfseries 283} (2023) 108586}
  [\href{https://arxiv.org/abs/2102.01031}{{\ttfamily 2102.01031}}].

\bibitem{Gonzalez:2022mcx}
D.~Gonzalez, N.~Kitajima, F.~Takahashi and W.~Yin, \emph{{Stability of domain
  wall network with initial inflationary fluctuations and its implications for
  cosmic birefringence}},
  \href{https://doi.org/10.1016/j.physletb.2023.137990}{\emph{Phys. Lett. B}
  {\bfseries 843} (2023) 137990}
  [\href{https://arxiv.org/abs/2211.06849}{{\ttfamily 2211.06849}}].

\bibitem{Kitajima:2023kzu}
N.~Kitajima, J.~Lee, F.~Takahashi and W.~Yin, \emph{{Stability of domain walls
  with inflationary fluctuations under potential bias, and gravitational wave
  signatures}},  \href{https://arxiv.org/abs/2311.14590}{{\ttfamily
  2311.14590}}.

\bibitem{Kitajima:2022jzz}
N.~Kitajima, F.~Kozai, F.~Takahashi and W.~Yin, \emph{{Power spectrum of
  domain-wall network, and its implications for isotropic and anisotropic
  cosmic birefringence}},
  \href{https://doi.org/10.1088/1475-7516/2022/10/043}{\emph{JCAP} {\bfseries
  10} (2022) 043} [\href{https://arxiv.org/abs/2205.05083}{{\ttfamily
  2205.05083}}].

\bibitem{Planck:2018vyg}
{\scshape Planck} collaboration, N.~Aghanim et~al., \emph{{Planck 2018 results.
  VI. Cosmological parameters}},
  \href{https://doi.org/10.1051/0004-6361/201833910}{\emph{Astron. Astrophys.}
  {\bfseries 641} (2020) A6}
  [\href{https://arxiv.org/abs/1807.06209}{{\ttfamily 1807.06209}}].

\bibitem{Babichev:2021uvl}
E.~Babichev, D.~Gorbunov, S.~Ramazanov and A.~Vikman, \emph{{Gravitational
  shine of dark domain walls}},
  \href{https://doi.org/10.1088/1475-7516/2022/04/028}{\emph{JCAP} {\bfseries
  04} (2022) 028} [\href{https://arxiv.org/abs/2112.12608}{{\ttfamily
  2112.12608}}].

\bibitem{Babichev:2023pbf}
E.~Babichev, D.~Gorbunov, S.~Ramazanov, R.~Samanta and A.~Vikman,
  \emph{{NANOGrav spectral index \ensuremath{\gamma}=3 from melting domain
  walls}}, \href{https://doi.org/10.1103/PhysRevD.108.123529}{\emph{Phys. Rev.
  D} {\bfseries 108} (2023) 123529}
  [\href{https://arxiv.org/abs/2307.04582}{{\ttfamily 2307.04582}}].

\bibitem{Dankovsky:2024ipq}
I.~Dankovsky, S.~Ramazanov, E.~Babichev, D.~Gorbunov and A.~Vikman,
  \emph{{Numerical analysis of melting domain walls and their gravitational
  waves}},  \href{https://arxiv.org/abs/2410.21971}{{\ttfamily 2410.21971}}.

\bibitem{Emond:2021vts}
W.~T. Emond, S.~Ramazanov and R.~Samanta, \emph{{Gravitational waves from
  melting cosmic strings}},
  \href{https://doi.org/10.1088/1475-7516/2022/01/057}{\emph{JCAP} {\bfseries
  01} (2022) 057} [\href{https://arxiv.org/abs/2108.05377}{{\ttfamily
  2108.05377}}].

\bibitem{Tkachev:1995md}
I.~I. Tkachev, \emph{{Phase transitions at preheating}},
  \href{https://doi.org/10.1016/0370-2693(96)00297-3}{\emph{Phys. Lett. B}
  {\bfseries 376} (1996) 35}
  [\href{https://arxiv.org/abs/hep-th/9510146}{{\ttfamily hep-th/9510146}}].

\bibitem{Greene:1997fu}
P.~B. Greene, L.~Kofman, A.~D. Linde and A.~A. Starobinsky, \emph{{Structure of
  resonance in preheating after inflation}},
  \href{https://doi.org/10.1103/PhysRevD.56.6175}{\emph{Phys. Rev. D}
  {\bfseries 56} (1997) 6175}
  [\href{https://arxiv.org/abs/hep-ph/9705347}{{\ttfamily hep-ph/9705347}}].

\bibitem{Daido:2017wwb}
R.~Daido, F.~Takahashi and W.~Yin, \emph{{The ALP miracle: unified inflaton and
  dark matter}},
  \href{https://doi.org/10.1088/1475-7516/2017/05/044}{\emph{JCAP} {\bfseries
  05} (2017) 044} [\href{https://arxiv.org/abs/1702.03284}{{\ttfamily
  1702.03284}}].

\bibitem{Daido:2017tbr}
R.~Daido, F.~Takahashi and W.~Yin, \emph{{The ALP miracle revisited}},
  \href{https://doi.org/10.1007/JHEP02(2018)104}{\emph{JHEP} {\bfseries 02}
  (2018) 104} [\href{https://arxiv.org/abs/1710.11107}{{\ttfamily
  1710.11107}}].

\bibitem{Co:2017mop}
R.~T. Co, L.~J. Hall and K.~Harigaya, \emph{{QCD Axion Dark Matter with a Small
  Decay Constant}},
  \href{https://doi.org/10.1103/PhysRevLett.120.211602}{\emph{Phys. Rev. Lett.}
  {\bfseries 120} (2018) 211602}
  [\href{https://arxiv.org/abs/1711.10486}{{\ttfamily 1711.10486}}].

\bibitem{Harigaya:2019qnl}
K.~Harigaya and J.~M. Leedom, \emph{{QCD Axion Dark Matter from a Late Time
  Phase Transition}},
  \href{https://doi.org/10.1007/JHEP06(2020)034}{\emph{JHEP} {\bfseries 06}
  (2020) 034} [\href{https://arxiv.org/abs/1910.04163}{{\ttfamily
  1910.04163}}].

\bibitem{Cline:1998rc}
J.~M. Cline, J.~R. Espinosa, G.~D. Moore and A.~Riotto, \emph{{String mediated
  electroweak baryogenesis: A Critical analysis}},
  \href{https://doi.org/10.1103/PhysRevD.59.065014}{\emph{Phys. Rev. D}
  {\bfseries 59} (1999) 065014}
  [\href{https://arxiv.org/abs/hep-ph/9810261}{{\ttfamily hep-ph/9810261}}].

\bibitem{Agrawal:2019lkr}
P.~Agrawal, A.~Hook and J.~Huang, \emph{{A CMB Millikan experiment with cosmic
  axiverse strings}},
  \href{https://doi.org/10.1007/JHEP07(2020)138}{\emph{JHEP} {\bfseries 07}
  (2020) 138} [\href{https://arxiv.org/abs/1912.02823}{{\ttfamily
  1912.02823}}].

\bibitem{Takahashi:2020tqv}
F.~Takahashi and W.~Yin, \emph{{Kilobyte Cosmic Birefringence from ALP Domain
  Walls}}, \href{https://doi.org/10.1088/1475-7516/2021/04/007}{\emph{JCAP}
  {\bfseries 04} (2021) 007}
  [\href{https://arxiv.org/abs/2012.11576}{{\ttfamily 2012.11576}}].

\bibitem{Yin:2021kmx}
W.~W. Yin, L.~Dai and S.~Ferraro, \emph{{Probing cosmic strings by
  reconstructing polarization rotation of the cosmic microwave background}},
  \href{https://doi.org/10.1088/1475-7516/2022/06/033}{\emph{JCAP} {\bfseries
  06} (2022) 033} [\href{https://arxiv.org/abs/2111.12741}{{\ttfamily
  2111.12741}}].

\bibitem{Jain:2022jrp}
M.~Jain, R.~Hagimoto, A.~J. Long and M.~A. Amin, \emph{{Searching for
  axion-like particles through CMB birefringence from string-wall networks}},
  \href{https://doi.org/10.1088/1475-7516/2022/10/090}{\emph{JCAP} {\bfseries
  10} (2022) 090} [\href{https://arxiv.org/abs/2208.08391}{{\ttfamily
  2208.08391}}].

\bibitem{Masia-Roig:2019hsy}
H.~Masia-Roig et~al., \emph{{Analysis method for detecting topological defect
  dark matter with a global magnetometer network}},
  \href{https://doi.org/10.1016/j.dark.2020.100494}{\emph{Phys. Dark Univ.}
  {\bfseries 28} (2020) 100494}
  [\href{https://arxiv.org/abs/1912.08727}{{\ttfamily 1912.08727}}].

\bibitem{GNOME:2023rpz}
{\scshape GNOME} collaboration, S.~Afach et~al., \emph{{What Can a GNOME Do?~
  Search Targets for the Global Network of Optical Magnetometers for Exotic
  Physics Searches}},
  \href{https://doi.org/10.1002/andp.202300083}{\emph{Annalen Phys.} {\bfseries
  2023} (2023) 2300083} [\href{https://arxiv.org/abs/2305.01785}{{\ttfamily
  2305.01785}}].

\end{thebibliography}\endgroup
\end{document}